%% file: ragan_tangents.tex
\documentclass[fleqn,usenatbib]{mnras}

\usepackage{newtxtext,newtxmath}
\usepackage{mathptmx}
\usepackage{txfonts}

\usepackage[T1]{fontenc}
\usepackage{ae,aecompl}

\usepackage{graphicx}	% Including figure files
\usepackage{amsmath}	% Advanced maths commands
\usepackage{amssymb}	% Extra maths symbols
\usepackage{caption}
\usepackage{soul}
\usepackage{rotating}
\usepackage{color}
\usepackage{graphics}
\usepackage{multirow}
\usepackage[T1]{fontenc}
\usepackage{ae,aecompl}
\usepackage{mathptmx}
\def\micron{\,$\mu$m\,}
\def\deg{$^{\circ}$}
\def\aap{A\&A}
\def\apj{ApJ}
\def\apjl{ApJL}
\def\apjs{ApJS}
\def\mnras{MNRAS}

\def\aj{AJ}
\def\pasp{PASP}
\def\nat{Nature}
\def\hone{H{\small I}}
\def\htwo{H{\small II}}
\def\gs{\mathrel{\raise1.5pt\hbox{$>$}\kern-6.0pt
\lower2.5pt\hbox{{$\sim$}}}}
\def\ls{\mathrel{\raise1.5pt\hbox{$<$}\kern-6.0pt 
\lower2.5pt\hbox{{$\sim$}}}}
\title[Star formation in spiral arms]{The role of spiral arms in Milky Way star formation}

\author[S.\,E.\,~Ragan et al.]{S.\,E. Ragan$^{1,2}$\thanks{email: RaganSE@cardiff.ac.uk}, T.\,J.\,T. Moore$^3$, D.\,J. Eden$^3$, M.\,G. Hoare$^2$, J.\,S.\, Urquhart$^4$, \and D. Elia$^5$, S. Molinari$^5$\\
\\
$^1$ School of Physics and Astronomy, Cardiff University, Queen's Buildings, The Parade, Cardiff, CF24 3AA, UK \\
$^2$ School of Physics and Astronomy, University of Leeds, EC Stoner Building, Leeds, LS2 9JT, UK \\
$^3$ Astrophysics Research Institute, Liverpool John Moores University, IC2, Liverpool Science Park, 146 Brownlow Hill, Liverpool, L3 5RF, UK \\
$^4$ School of Physical Sciences, Ingram Building, University of Kent, Canterbury, Kent CT2 7NH, UK \\
$^5$ INAF-Istituto di Astrofisica e Planetologia Spaziali, Via Fosso del Cavaliere 100, I-00133 Roma, Italy 
}

\date{Accepted 2018 June 20. Received 2018 June 20; in original form 2017 October 19}

\pubyear{2018}

\begin{document}
\label{firstpage}
\pagerange{\pageref{firstpage}--\pageref{lastpage}}
\maketitle

\begin{abstract}
What role does Galactic structure play in star formation? We have used the {\em Herschel} Hi-GAL compact-clump catalogue to examine trends in evolutionary stage over large spatial scales in the inner Galaxy. We examine the relationship between the fraction of clumps with embedded star formation (the star-forming fraction, or SFF) and other measures of star-formation activity. Based on a positive correlation between SFF and evolutionary indicators such as the luminosity-to-mass ratio, we assert that the SFF principally traces the average evolutionary state of a sample and must depend on the local fraction of rapidly-evolving, high-mass young stellar objects. The spiral-arm tangent point longitudes show small excesses in the SFF, though these can be accounted for by a small number of the most massive clusters, just 7.6\% of the total number of clumps in the catalogue. This suggests that while the arms tend to be home to the Galaxy's massive clusters, the remaining 92.4\% of Hi-GAL clumps in our catalogue do not show an enhancement of star formation within arms. Globally, the SFF is highest at the Galactic midplane and inner longitudes. We find no significant trend in evolutionary stage as a function of position across spiral arms at the tangent-point longitudes. This indicates that the angular offset observed between gas and stars, if coordinated by a density wave, is not evident at the clump phase; alternatively, the onset of star formation is not triggered by the spiral density wave. 
\end{abstract}


\section{Introduction}

Spiral galaxies organise most of their molecular gas in coherent arm structures, but the origin and persistence of spiral structure and the associated star formation has been a topic of debate for decades. The theoretical debate largely looms over the longevity of spiral arms. Density wave theory \citep{Lindblad1960,LinShu1964} posits that spiral arms are due to quasi-steady (long-lived) global modes of the disc. As a spiral density wave moves at its pattern speed through the disc, gas falls into the minimum of the spiral potential, is compressed into molecular clouds and forms stars. \citet{Roberts1969} argue that star formation is directly triggered and this should manifest itself in spatial offsets between gas and tracers of star formation. 

A long-standing challenge to classical density wave theory has been that the long-lived spiral modes could not be sustained naturally in the disc \citep{Toomre1969}. A growing number of galaxy simulations have led to the emergence of an alternative picture where spiral structure can develop dynamically, without the imposition of a fixed potential. In this sort of scenario, arms tend to be short-lived, transient structures that result from recurring gravitational instabilities. Star formation is thus stochastic in nature and would not exhibit any spatially ordered pattern. For a full review of numerical simulations of spiral structure, we refer the reader to \citet{DobbsBaba2014}.

Spatial offsets between gas and star formation tracers have been sought observationally in nearby spiral galaxies \citep[e.g.][]{Tamburro2008,Egusa2009,Foyle2011} with some discrepancy in the interpretations, due at least in part to the different choice of tracers \citep[see][]{Louie2013} and large empirical scatter in how gas and stars are  related \citep{Schinnerer2013}. To avoid these difficulties, some authors have looked at variations within the properties of a single type of tracer. For example,
\cite{Choi2015} analysed resolved stellar populations to determine the star-formation history as a function of position across the arms of the grand-design spiral M81 but found no evidence of the propagating star formation across the arm that is the prediction of the density-wave model. \citet{Tenjes2017} find no systematic offsets between ultraviolet, infrared and CO in M31. \citet{Schinnerer2017} examine the ISM properties across an arm in grand-design spiral, M51, and find no significant variations \citep[cf.][]{Vogel1988}.

Another way to investigate the impact of spiral arms is to compare the star formation in arms versus interarm regions in galaxies of different morphologies. \citet{Foyle2010} find no enhancement in star formation efficiency (SFE) in the spiral-arm regions compared to interarm regions in their small sample of nearby galaxies, suggesting that the SFR per unit gas mass is not enhanced in arms. As this idea is investigated in more galaxies \citep[cf.][]{Rebolledo2015, Kreckel2016}, there are some instances of apparent SFE enhancement in arms, though samples sizes are too small to draw general conclusions about whether certain galaxy properties (e.g. morphology) correlate with this phenomenon. 

In the Milky Way, \citet{HeyerTerebey1998} interpret the increase in the H$_2$/H{\small I} fraction in the Perseus spiral arm compared to interarm gas as an indication of increased cloud formation efficiency in the arm and thus evidence of a triggered enhancement of the SFE. \citet{Sawada2012a,Sawada2012b} argue that brighter and more compact CO emission in arms compared to interam regions indicates that the arm triggers small-scale collapse of molecular gas structures. These observed phenomena could either be the effect of molecular clouds being longer-lived within spiral arms \citep{Elmegreen1986,Roman-Duval2010,Dobbs2011} or the result of statistics, with more extreme clouds tending to be found in the larger samples from the more densely populated arms, or some combination of both. 

Within molecular clouds, \citet{Eden2012, Eden2013} find no difference in the fraction of molecular gas found in dense clumps when comparing samples from the arm and interarm regions in the inner Galaxy and, likewise, for the fraction of dense clumps exhibiting signs of star formation \citep{Eden2015}. The main contributor to any spatial variations in these measures is cloud-to-cloud scatter, with extreme star-forming complexes being the largest outliers affecting averages on kiloparsec-scales \citep[see also][]{Moore2012}. 

The above threads of investigation have provided conflicting information and thus we have no consensus about the impact that spiral arms have on star formation in the Galaxy. The picture is further complicated by the fact that star formation throughout the Milky Way is patchy in space \citep[e.g.][]{Urquhart2017} and therefore star formation is likely to be intermittent in time as well. Moreover, the effect of the Galactic bar on star formation within 3-4\,kpc of the Galactic Centre is poorly understood, but the presence of a bar may suppress star formation \citep{James2016} and (as simulations have shown) disrupt the dynamics throughout the disc \citep[e.g.][]{Dobbs2010}. 

The \textit{Herschel} Hi-GAL survey provides a much more detailed look at the properties of star-forming clumps throughout the Galaxy than any previous survey. \citet{Elia2017} have produced a compact-source catalogue containing $>$10$^5$ clumps within the inner Galaxy and, by modelling their physical properties, classified them based on their evolutionary stage. In an initial study \citep{Ragan2016}, we investigated the prevalence of star formation (as measured by the fraction of clumps in a given area harbouring embedded star formation, the so-called ``star-forming fraction'' or SFF) as a function of Galactocentric radius ($R_\mathrm{GC}$) and found that, while there is a gradual but statistically significant decline in SFF with $R_\mathrm{GC}$ in the inner disc (3\,kpc $< R_\mathrm{GC} <$ 8\,kpc), radii associated with spiral arms do not stand out in the SFF versus $R_\mathrm{GC}$ plane. In this paper, we use the SFF parameter to look for spatial trends in evolutionary stage associated with the Galactic spiral arms, which could be analogous to those sometimes reported in nearby galaxies.

%%%%%%%%%%%%%%%%%%%%%
\section{Data}

\subsection{Hi-GAL}

The {\it Herschel} key program Hi-GAL \citep{Molinari2010a,Molinari2010b} surveyed the Galactic plane in the five bands available with the PACS \citep[70 and 160\micron;][]{A&ASpecialIssue-PACS} and SPIRE \citep[250, 350 and 500\micron;][]{A&ASpecialIssue-SPIRE} instruments. These wavelengths cover the peak of the spectral energy distribution of thermal emission from dust grains in the temperature range 8\,K $< T_{\rm dust} <$ 50\,K. Compact sources at these wavelengths have the cold, dense conditions necessary for star formation.

We use the Hi-GAL compact-source catalogue \citep{Elia2017}, which contains over 10$^5$ sources within the inner Galactic longitudes $-71^{\circ} \le \ell \le +67^{\circ}$, detected in at least three adjacent bands -- either 160, 250 and 350\micron or 250, 350 and 500\micron. Most Hi-GAL sources do not have a reliable distance estimate but, for a subset, a rotation-curve-based method described in \citet{Russeil2011} was used to assign velocities and distances to $\sim$56\% of the sources. Encouragingly, \citet{Ragan2016} showed that, even considering the generally large uncertainties associated with kinematic distance estimates, assuming peculiar velocities are isotropic, the large-scale trends in SFF are robust. Nevertheless, the inherent limitation to distance estimation is a strong motivation to focus our study on tangent-point longitudes, where physical distance from the arm leading edge translates to a longitudinal offset. In forthcoming work (Russeil et al., in preparation), a more refined distance estimation method will be applied to the entire Hi-GAL survey.

\subsection{Spiral arm model}

The Milky Way is apparently a four-arm, barred, trailing-arm spiral galaxy \citep{BinneyTremaine}. Two of these arms -- Scutum-Centaurus and Perseus -- are considered to be the ``dominant arms'', based on their strength in both gas and stellar tracers, while the other two arms -- Norma and Sagittarius -- are weaker features, especially with respect to high-mass stars \citep{Robitaille2012}. In addition to the four arms, the so-called expanding 3\,kpc arm is a dominant feature in the gas distribution on the near and far side of the Galactic centre region \citep[see][and references therein]{Dame2008}. The 3\,kpc arm(s) are believed to originate at the end(s) of the Galactic bar. 

The precise path of the arms in position and velocity space is uncertain and subject to a considerable amount of on-going work \citep[cf.][]{Vallee1995,HouHan2014,Reid2016}. In addition to the uncertainty of the spiral-arm morphologies, it is well-known that different tracers of spiral arms -- CO emission \citep[e.g.][]{Roman-Duval2009}, \htwo\  regions \citep[e.g.][]{Urquhart2012}, methanol masers \citep[e.g.][]{Green2017}, dust emission \citep[e.g.][]{Beuther2012} and stars \citep[e.g.][]{Benjamin2005} -- are offset from one another \citep{Vallee2014a,HouHan2015,Vallee2016b} when measured at the arm tangent points. For the purposes of this paper, we adopt the tangent-point positions from \citet{HouHan2015}, which are summarised in Table \ref{t:tangentlong}. In their paper, \citet{HouHan2015} examine the longitudinal distributions of several gas and dust tracers, as well as tracers of the ``old'' stellar populations. We adopt the peak longitudes of the ATLASGAL dust source distribution as such sources are the closest analogues in the \citet{HouHan2015} study to the compact Hi-GAL sources that we use in this paper. 

To determine the width each arm subtends in longitude, we use the model presented in \citet{Reid2014a} for arm width (in parsecs) as a function of $R_\mathrm{GC}$. Using the approximate heliocentric distance from the Sun to the tangent point, we compute the corresponding angular width. The arm properties are summarise in Table \ref{t:tangentlong}. 

\input{houhan_tangents_props_tab.tex}

%%%%%%%%%%%%%%%%%%%%%%%%%%%%%%%%%%%
\section{Measures of star formation}

\subsection{Quantifying evolutionary stages}

A wide range of methods is employed to measure the amount of star formation occurring on different scales \citep{KennicuttEvans_ARAA}. Counting the number of young stellar objects (YSOs) is considered the ``gold standard'' for estimating the recent star forming activity \citep{Heiderman2010,Lada2010,Lombardi2013,Evans2014}, though the use of this method is only feasible in nearby clouds. On galactic scales, we rely on other tracers such as line emission from ionised gas, infrared or ultraviolet emission to infer the star formation rate surface density, though the calibration of these measures is much less certain and not even entirely consistent from one  tracer to another \citep{Vuti2013}. 

The advent of large-area Galactic-plane surveys has enabled us to study the statistical properties of star-forming clumps over a large volume of the Milky Way. The early phases of star formation are found in cold, dense clumps of gas and dust, where the latter emits largely in the far-infrared and sub-millimetre wavelength regimes. These clumps are the densest condensations within giant molecular clouds (GMCs) and so provide information on star formation within clouds as a function of location. One quantity of particular interest is the fraction of gas within a GMC at high densities, or the so-called dense gas mass fraction (DGMF). The DGMF is analogous to or a precursor to the SFE in clouds \citep{Eden2013} as it represents the first step in the conversion of molecular clouds to stars. The mean DGMF in the Milky Way is a few per cent \citep{Battisti2014} and does not vary significantly as a function of position in the inner Galactic disc when averaged over samples of clouds \citep{Ragan2016}, though large variations are seen from cloud to cloud \citep{Eden2012}.

Using Hi-GAL data, \citet{Elia2017} have established a framework for the evolutionary classification of individual compact clumps based on their spectral energy distributions (SEDs): infrared colour, bolometric dust temperature ($T_\mathrm{bol}$, defined as the temperature of a blackbody that has the same mean frequency as the SED, \citealp{Myers1993}), luminosity-to-mass ($L_\mathrm{bol}$ / $M_\mathrm{tot}$) ratio \citep[e.g.][]{Molinari2008,Urquhart2014a,Molinari2016c,Urquhart2017}, and $L_\mathrm{bol}$ / $L_\mathrm{submm}$ ratio \citep{Andre1993} diagnostics, all of which have the distinct advantage of being distance-independent. These quantities are particularly useful in comparing relative properties of individual clumps and YSOs, but integrated over kiloparsec scales, they are dominated by a small number of luminous sources \citep{Moore2012,Urquhart2017}. 

In an attempt to overcome the bias toward the most luminous sources, \citet{Ragan2016} defined a simple diagnostic called the star-forming fraction (SFF), which is the fraction of Hi-GAL sources in a given area that are 70-\micron bright\footnote{\citet{Ragan2016} defined 70-\micron-``bright'' as having 70-\micron flux density above a fixed threshold set by the $\sim$uniform diffuse background characterised in \citep{Molinari2016c}.}. This criterion, which factors heavily into the above-mentioned classification scheme, hinges on the fact that the presence of a 70-\micron counterpart is a reliable indication that a compact source contains active star formation \citep{Dunham2008,Ragan2012b}. By considering subsets of equidistant Hi-GAL clumps, we also showed that the detected spatial variations of SFF are robust against the varying luminosity sensitivity due to distance.

The SFF is potentially related to both the mean evolutionary stage, and so the time gradient of the star-formation rate, and to the efficiency of star formation within the dense clumps in a given area.  Thus, a higher SFF is indicative of more advanced stages or more efficient conversion of cold, dense material into stars and clusters compared to regions with lower SFF. This is also true of the $L_\mathrm{bol}$-based parameters, such as $L_\mathrm{bol}$ / $M_\mathrm{tot}$, but, even though the latter metric is not a strong function of clump mass (e.g., \citealp{Urquhart2017}), there are potential complications in its interpretation arising from sample-selection effects (see below) and the underlying non-linearity of the stellar mass-luminosity relation, while SFF simply counts the dense clumps that have evidence of star formation.

\begin{figure}
\centering
\includegraphics[width=0.45\textwidth]{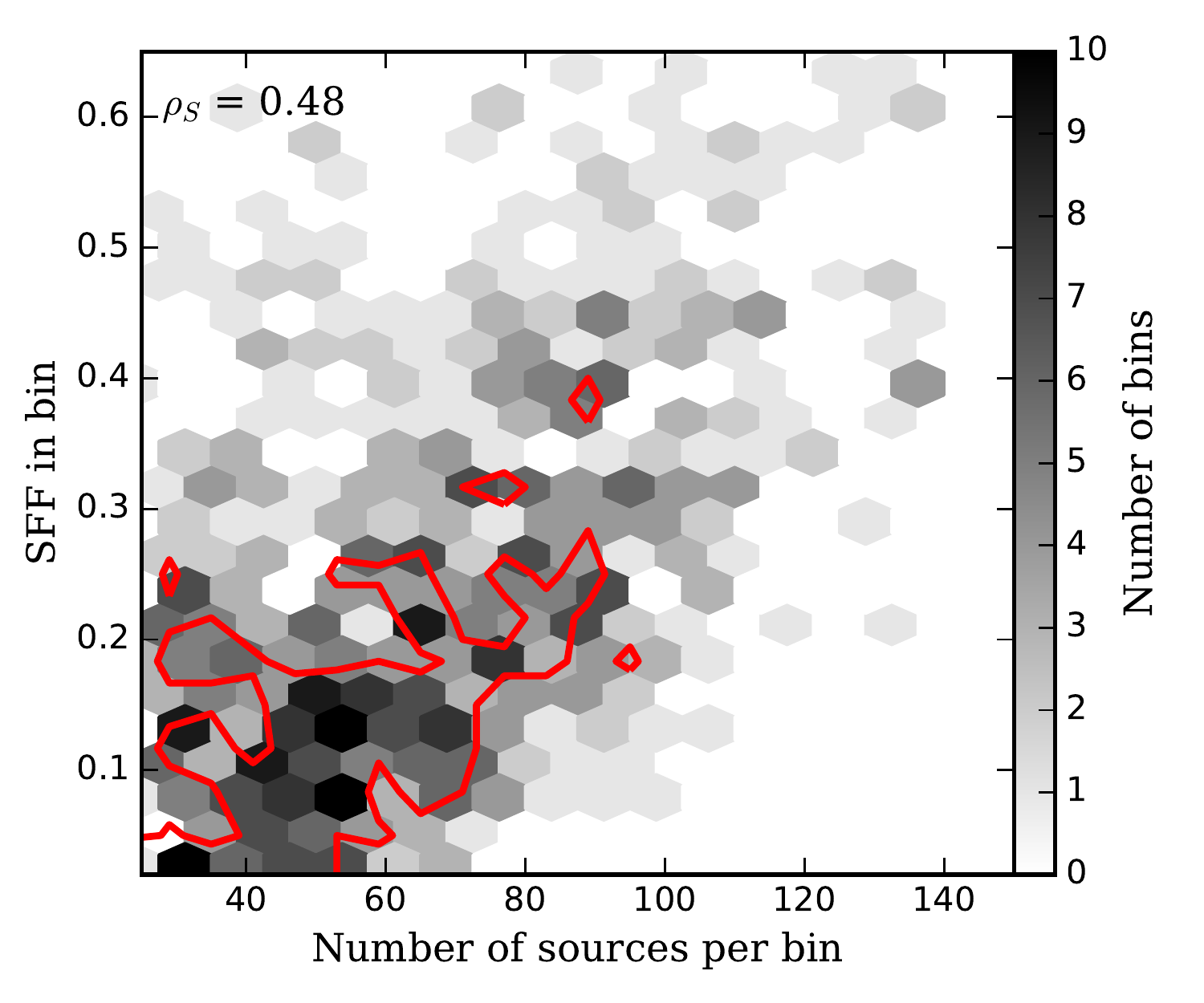}
\caption{SFF versus the number of sources per area bin, showing the dependence on sample size.  The greyscale indicates the number of bins having the same $N$ and SFF. The red contour encompasses bins with 5 or more counts. The relation shows a significant ($p < 0.0001$) positive correlation with a Spearman rank ($\rho_\mathrm{S}$ = 0.5).
\label{f:SFF_Nsrc}}
\end{figure}

\begin{figure}
\includegraphics[width=0.46\textwidth]{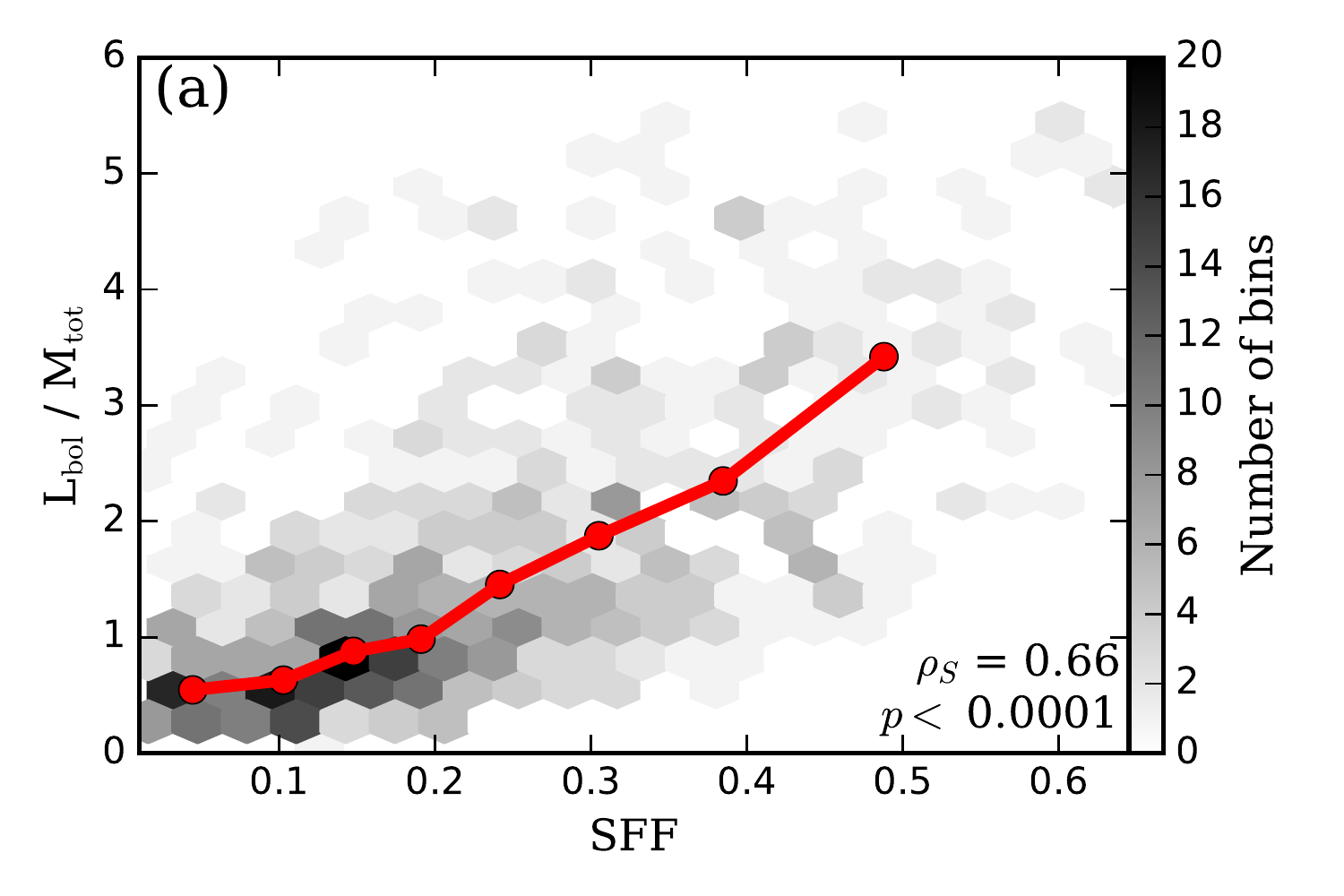} \
\includegraphics[width=0.46\textwidth]{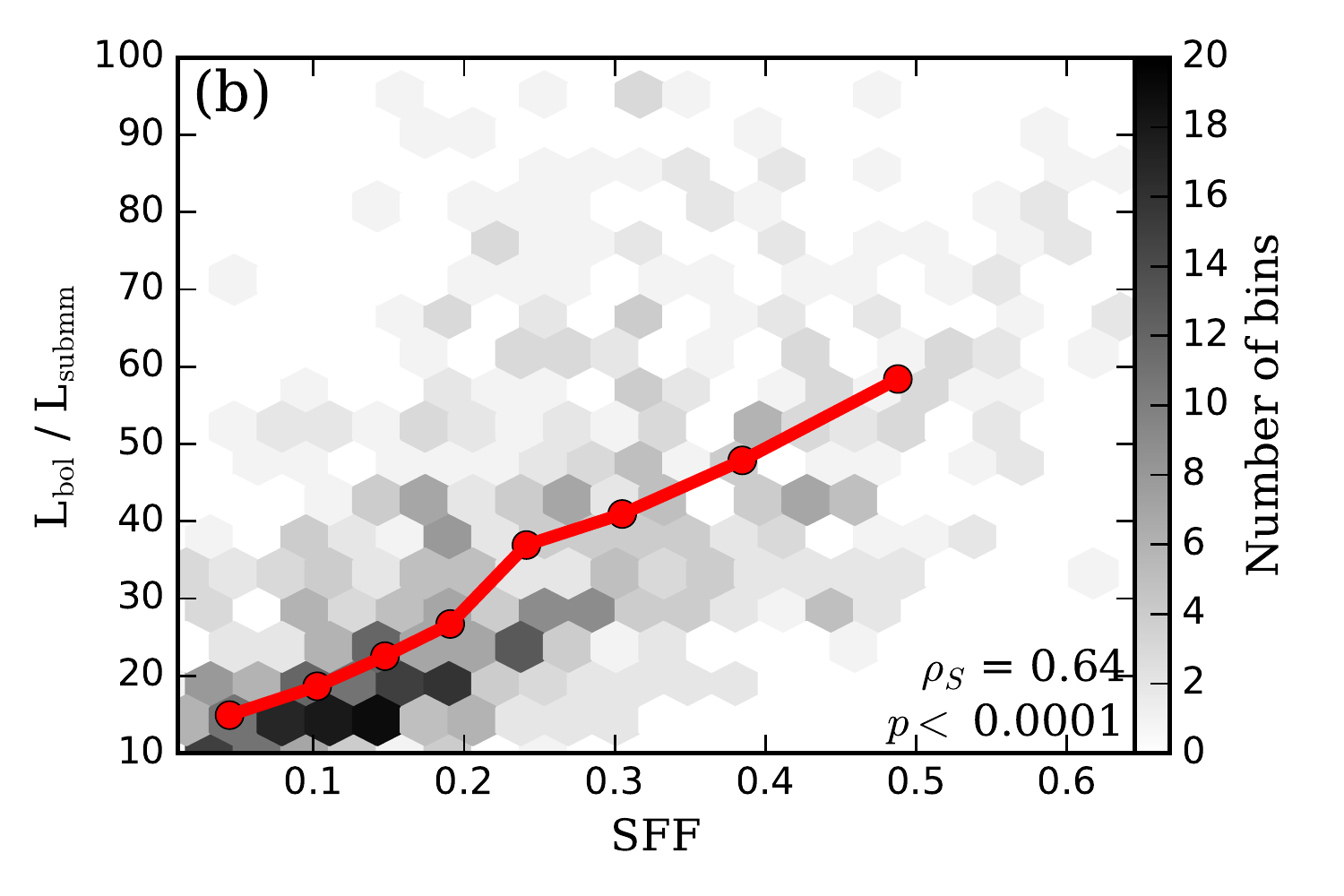} \
\includegraphics[width=0.46\textwidth]{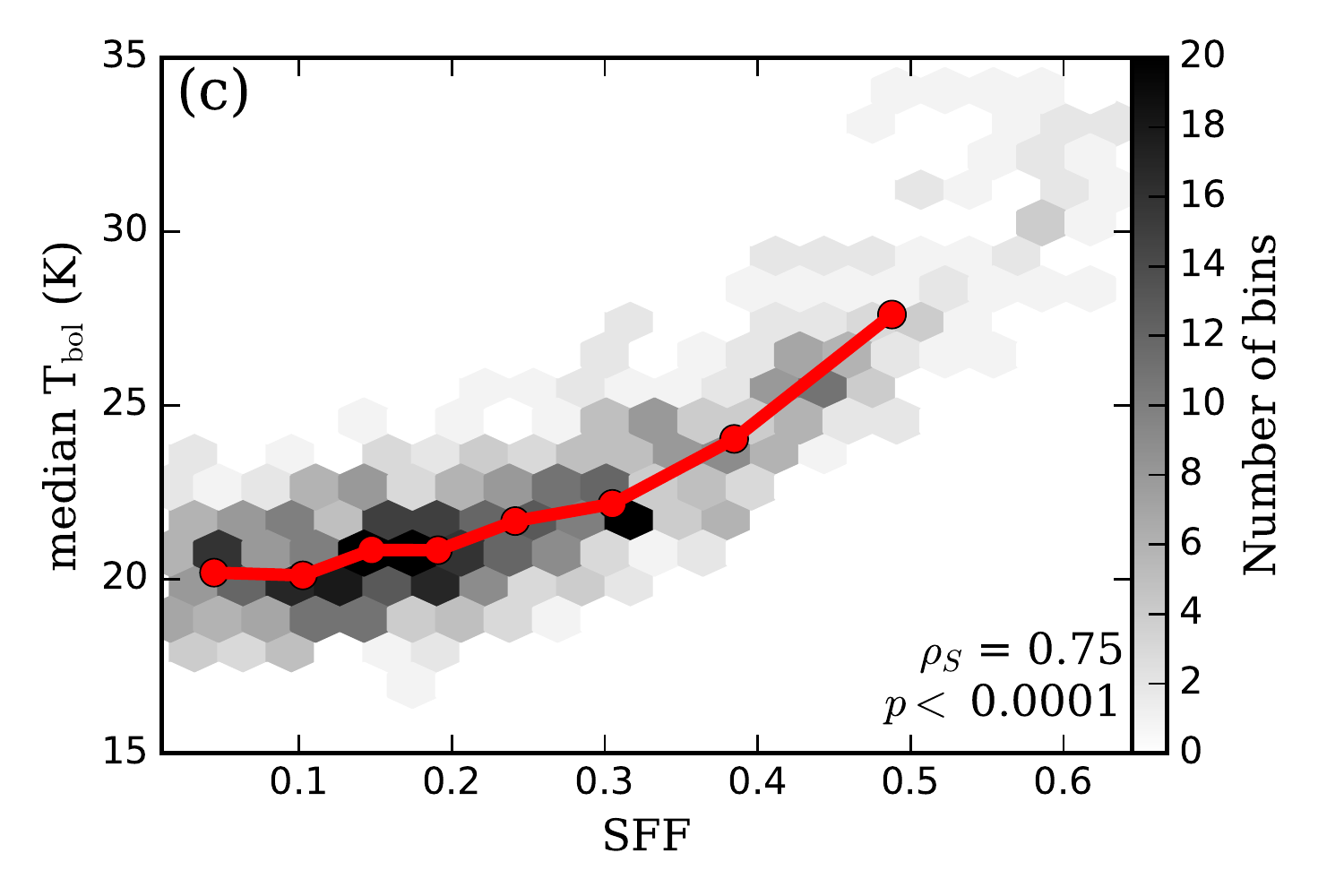}
\caption{(a) Plot of the $L_\mathrm{bol}$ / $M_\mathrm{tot}$ (in units of $L_{\odot}$ / $M_{\odot}$) as a function of the star-forming fraction (SFF). Each hexagon represents a two-dimensional bin (i.e. from Figures~\ref{f:SFFmap1} and \ref{f:SFFmap4}). (b) $L_\mathrm{bol}$ / $L_\mathrm{submm}$ versus SFF with similar colour encoding as above. (c) Median bolometric temperature ($T_\mathrm{bol}$) of all clumps in each bin. In all cases, the red line shows the median value computed over eight bins spaced such that they contain an equal number of points. The Spearman rank coefficient ($\rho_\mathrm{S}$) and $p$-value for each relation is shown in the lower right corner of each panel. \label{f:SFFvsLM}}
\end{figure}

\subsection{Mapping star formation}

We present maps in Galactic latitude versus longitude of total Hi-GAL compact source counts \citep{Elia2017} and compare with maps of the SFF, luminosity-to-mass and $L_\mathrm{bol}$ / $L_\mathrm{submm}$ ratios, calculated from the total luminosity and total mass in each bin in Figures~\ref{f:SFFmap1} and \ref{f:SFFmap4}. In all maps, we consider $|\,b\,| < $1\deg and 15\deg $< |\,\ell\,| <$ 65\deg, and the two-dimensional bins are 0.2\deg\ wide in latitude and 1.0\deg\ wide in longitude and are the same in all panels. The top panels displaying total source counts show that there is a great proportion of compact sources at smaller absolute longitudes; the inner half of the longitude range considered (15\deg $< |\,\ell\,| <$ 40\deg) contains 63.2\% of all sources. The vertical lines indicate the spiral-arm tangent-point longitudes in this range (see Table \ref{t:tangentlong}). 

The (b) panels of Figures~\ref{f:SFFmap1} and \ref{f:SFFmap4} show maps of the SFF, which ranges from 0.02 to 0.6 with a mean of 0.25. There are a greater number of localised peaks in SFF in the inner longitudes (15\deg $< |\,\ell\,| <$ 40\deg), consistent with the elevated SFF found at small Galactocentric radii corresponding to this longitude range \citep{Ragan2016}. Because of shorter evolution timescales in high-mass YSOs and a more rapid transition to the IR-bright stage \citep{Urquhart2014c}, there is a potential bias related to sample size, since larger samples will include more high-mass sources. In order to investigate this, we show, in Figure \ref{f:SFF_Nsrc}, the relation between SFF and the number of sources per two-dimensional bin. The plot shows a positive trend of SFF as a function of bin population with a Spearman-rank correlation coefficient ($\rho_\mathrm{S}$) of 0.5, giving a $p$-value $<$ 0.0001. This bias requires correction and so, in the following study of spatial trends, we use constant population bins.

The lower two panels, (c) and (d), of Figures~\ref{f:SFFmap1} and \ref{f:SFFmap4} reveal a close correspondence between the $L_\mathrm{bol}$/$M_\mathrm{tot}$ and $L_\mathrm{bol}$/$L_\mathrm{submm}$ ratios. This is unsurprising, since $M_\mathrm{tot}$ and $L_\mathrm{submm}$ are related via the dust temperature.
We show how these metrics relate to the SFF in Figure \ref{f:SFFvsLM}. There is a significant ($p<0.0001$) positive correlation between $L_\mathrm{bol}$/$M_\mathrm{tot}$ and SFF ($\rho_\mathrm{S}$ = 0.66), $L_\mathrm{bol}$/$L_\mathrm{submm}$ and SFF ($\rho_\mathrm{S}$ = 0.64) and the median $T_\mathrm{bol}$ and SFF ($\rho_\mathrm{S}$ = 0.75). The median values of these ratios as a function of SFF, distributed over eight bins containing an equal number of points is shown with the red lines in Figure \ref{f:SFFvsLM}. The scatter, particularly in panels (a) and (b), is largely due to the scatter and overlap in these properties between the populations of starless and protostellar clumps shown in \citet{Elia2017}. Nevertheless, the positive correlations between SFF and these metrics indicates that the SFF is indeed a measure of the prevalence of star formation.  SFF and both luminosity-based metrics are mixed parameters, tracing both SFE within clumps and mean evolutionary state and all three share the dependence on sample size. $T_\mathrm{bol}$ is a cleaner tracer of evolution and its tighter correlation with SFF shows that the latter also mainly traces the average evolutionary state of a sample.

Why not simply use familiar metrics such as $L_\mathrm{bol}$/$M_\mathrm{tot}$ or $T_\mathrm{bol}$ rather than invoking the SFF? While the common clump-based metrics to which we compare the SFF in Figure \ref{f:SFFvsLM} function well to ascertain \textit{relative} evolutionary stage between clumps, their aggregate values over large areas are dominated by a small number of extreme sources, where the SFF weights all clumps in a given area equally. \citet{Urquhart2017} have demonstrated that the thirty most massive complexes in the ATLASGAL survey account for $\sim$30\% of all dense gas in the inner Galaxy and $\sim$50\% of the total luminosity. Of those thirty complexes, 22 fall in the longitude range that we are considering in this work. The locations of these clusters are listed in Appendix \ref{a:atlasgal}, where we also tabulate the number of compact Hi-GAL sources encompassed and the overall SFF of those regions. In what follows, we will examine spatial trends in SFF and $L_\mathrm{bol}$/$M_\mathrm{tot}$ with latitude and longitude, both with and without this population of the most massive regions (henceforth referred to as ATLASGAL Massive Clusters, or AMCs), which encompass only 7.6\% of the Hi-GAL compact clumps by number. In doing so, we can investigate spatial trends in compact source properties without the dominating influence of the most massive regions. 

\section{Results}

\subsection{Latitude distribution of star formation}

\begin{figure}
\centering
\includegraphics[width=0.36\textwidth]{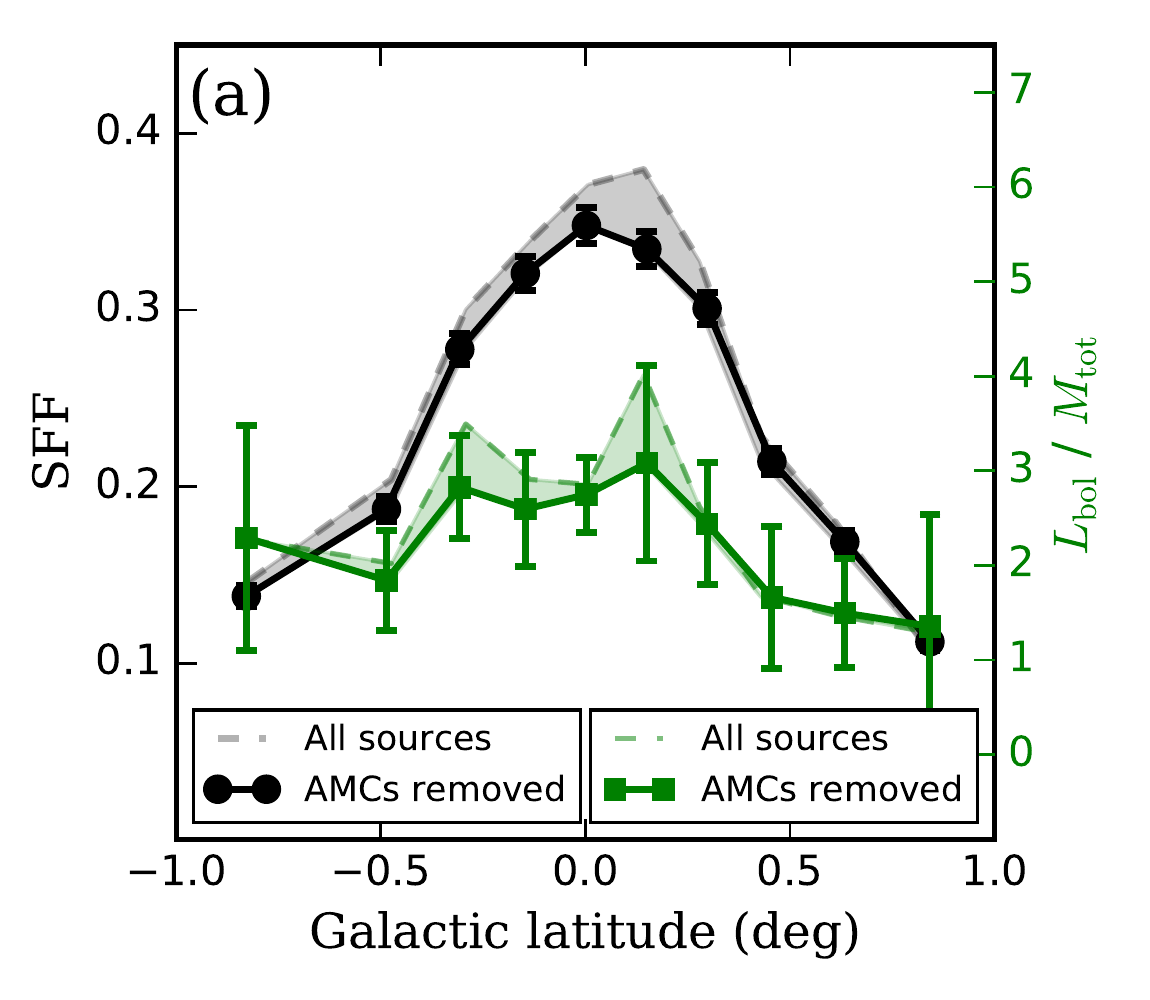} \
\includegraphics[width=0.36\textwidth]{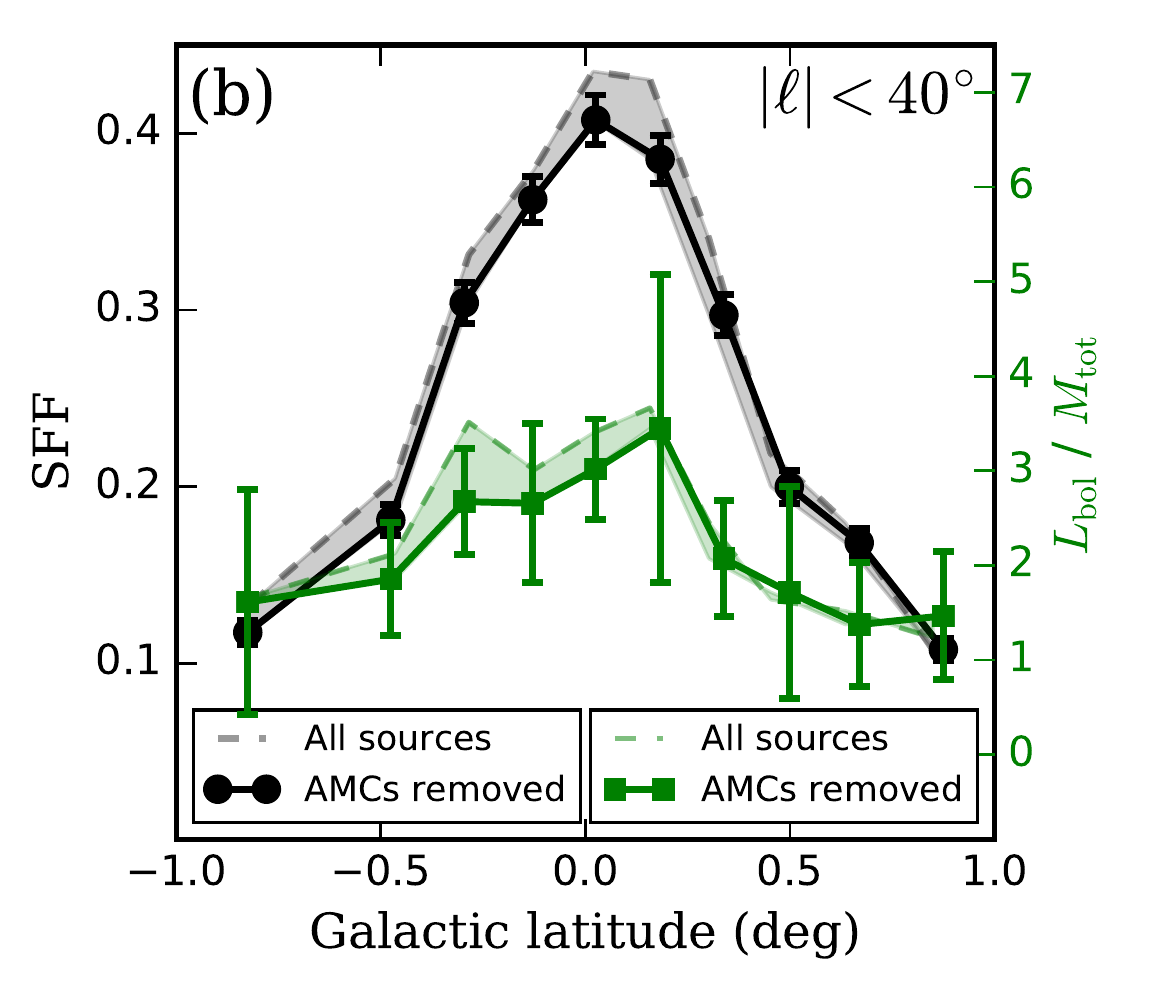} \
\includegraphics[width=0.36\textwidth]{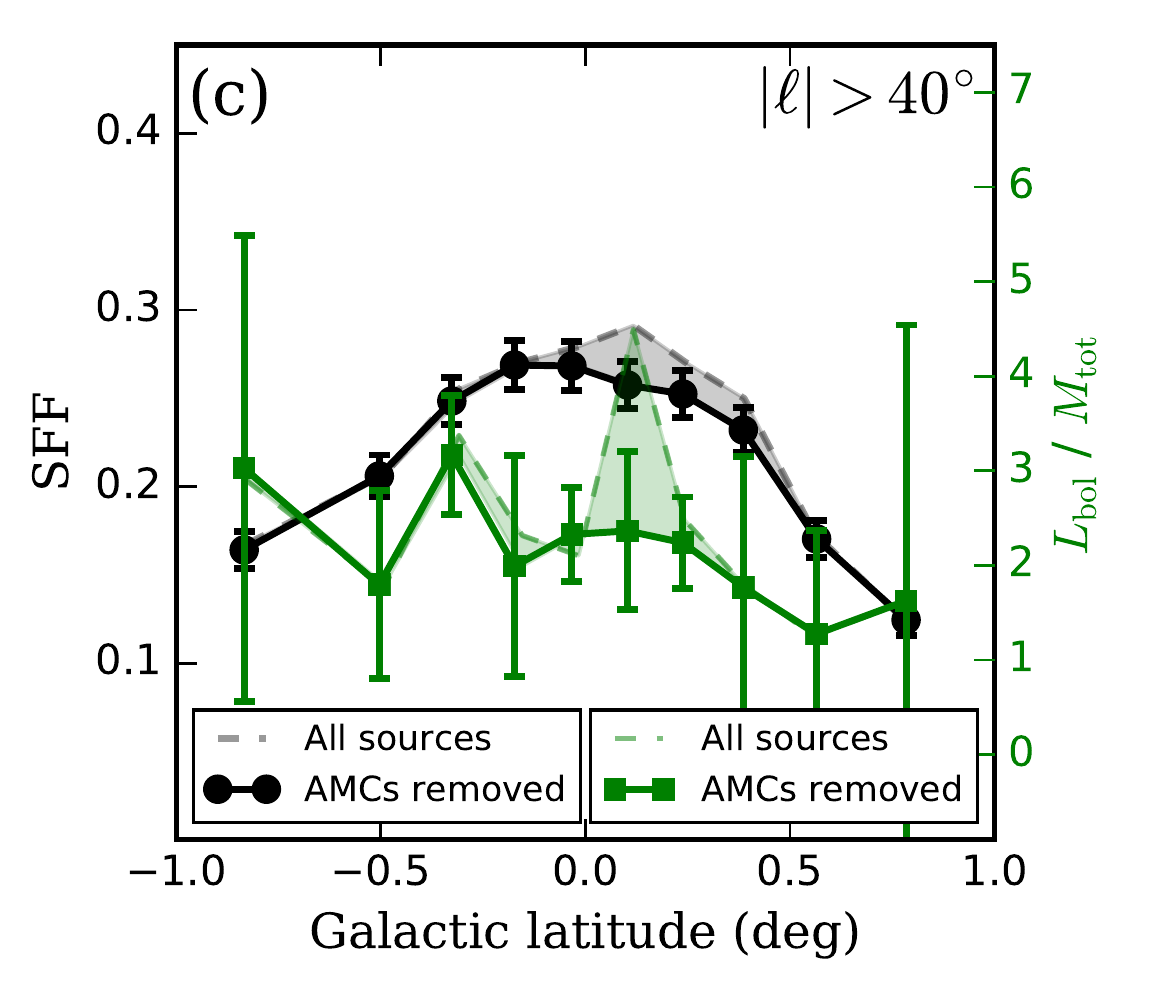}
\caption{Histograms of SFF and $L_\mathrm{bol}$ / $M_\mathrm{tot}$ (in units of $L_{\odot}$ / $M_{\odot}$) as a function of Galactic latitude in (a) all longitudes, (b) interior to $|\,\ell\,|$ = 40\deg and (c) exterior to $|\,\ell\,|$ = 40\deg. The ten bins in latitude are spaced so that they contain an equal number or sources. The light grey dashed curves show the distribution of SFF of all Hi-GAL clumps, and the black curves show the distribution after the removal of the ATLASGAL Massive Clusters (AMCs). The green dashed and solid curves (scale on the right axis) shows the $L_\mathrm{bol}$ / $M_\mathrm{tot}$ before and after AMC removal, respectively. The errorbars indicate the standard error of the mean.
\label{f:SFFvsGLAT}}
\end{figure}

In the inner Galaxy, dense clumps are narrowly confined to the midplane, as are nearly all of the  prodigious star forming regions \citep[][]{Urquhart2014a,Urquhart2017}. The sample of Hi-GAL clumps we study here enables us to explore the prevalence and evolutionary phase of star formation in a broader clump population. In Figure \ref{f:SFFvsGLAT}a, we plot the SFF and $L_\mathrm{bol}$/$M_\mathrm{tot}$ as a function of latitude, over bins spaced such that each contains an equal number of clumps. The SFF distribution is symmetric about the midplane, but the $L_\mathrm{bol}$/$M_\mathrm{tot}$ trend is less regular. We show the effect of removing AMCs from the sample in the solid curves. The shape of the SFF distribution changes little, while the $L_\mathrm{bol}$/$M_\mathrm{tot}$ flattens. In panels b and c of Figure \ref{f:SFFvsGLAT}, we examine the trends for the inner and outer halves of the considered longitude range separately, showing that for the inner ($|\,\ell\,| < 40$\deg) subset, the SFF is more strongly peaked at the midplane. 

That the SFF versus latitude trend is less affected by the removal of the AMCs than the $L_\mathrm{bol}$/$M_\mathrm{tot}$ over the same range goes back to the small \textit{number} of sources associated with the AMCs, thus leaving the SFF relatively unchanged. The large share of the luminosity and mass that the AMCs comprise, on the other hand, affects larger localised changes upon their removal. 

\begin{figure*}
\centering
\includegraphics[width=\textwidth]{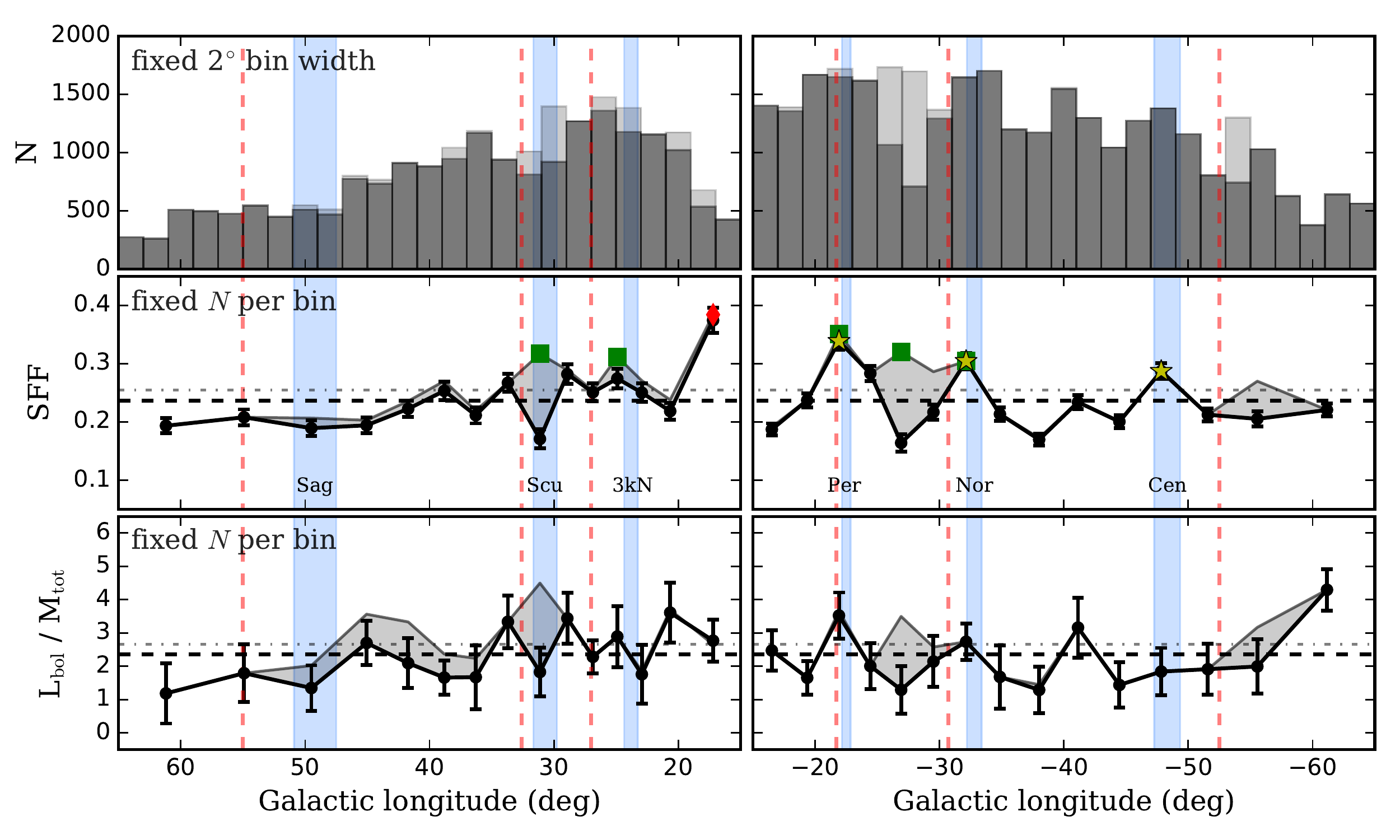}
\caption{In the top panels, we plot the number of sources in each 2-degree longitude bin for the first (left) and fourth (right) quadrants. The light grey histogram shows all Hi-GAL clumps, and the dark grey histogram shows the distribution after the removal of the AMCs. The middle panels show the SFF as a function of Galactic longitude. The grey curve shows the distribution of all Hi-GAL sources over the longitude interval with 15 bins spaced to contain equal number of sources. Green squares indicate longitudes where the SFF in a bin exceeds 3$\sigma$ above the mean SFF (0.254, shown in the grey dot-dashed line). The black curve shows the distribution after the AMCs have been removed. The revised mean SFF of the remaining clumps (0.233) is shown in the dashed black line. The yellow stars show the $>$3$\sigma$ peaks that remain after this editing. The ``peak'' indicated with the red diamond at the innermost bin is not necessarily a local maximum, despite having the highest SFF in the considered longitude range. The lower panels are the same except that we plot the $L_\mathrm{bol}$ / $M_\mathrm{tot}$ ratio in units of $L_{\odot}$ / $M_{\odot}$, with error bars showing the standard error of the mean. In all panels, the vertical blue shaded regions are the longitudes associated with the spiral arm tangents (see Table \ref{t:tangentlong}) and the vertical red dashed lines are median longitudes of the peaks in the stellar population \citep{HouHan2015}.
\label{f:SFF_glong}}
\end{figure*}

\subsection{Star formation as a function of longitude}

Searching for spatial variation in star formation in the Milky Way poses unique challenges compared to analogous studies of nearby galaxies. One of the major complications is confusion along the line of sight. An arbitrary direction toward the inner Galaxy will intersect with multiple spiral arms. Any spatial offsets between evolutionary stages (e.g., gas and stars) expected from density-wave theory are on the same order, if not smaller than typical heliocentric distance uncertainties using current (kinematic) methods. We therefore focus our attention on the tangent-point longitudes, where such confusion is minimised and we can assume that clumps on given lines of sight are overwhelmingly located at a similar position within the arm.

In Figure \ref{f:SFF_glong}, the top panels show the source-count distribution in equally-spaced, 2\deg-wide longitude bins. The blue vertical shaded areas show the tangent-point longitudes, the widths of which correspond to their estimated $R_\mathrm{GC}$-dependent width \citep{Reid2014a}. In the second and third rows of Figure \ref{f:SFF_glong}, we show the SFF and $L_\mathrm{bol}$/$M_\mathrm{tot}$ ratio, respectively, as a function of longitude. To compute these values, we summed over all Galactic latitudes and arranged the data into equal-population longitude bins. The distribution of the full sample is shown in the thin grey curve, and the black curve shows the distribution after the AMCs (see Table \ref{t:atlasgal}) are removed. 

\input{peak_longitudes.tex}

The distribution of SFF with longitude has six peaks (green squares in Figure \ref{f:SFF_glong}) that are $>$3$\sigma$ ($\sigma$ here being the standard error on the mean) above the mean value of 0.254 (grey dash-dotted line in Figure \ref{f:SFF_glong}) in the first and fourth quadrants before the removal of the AMCs. These are listed in Table \ref{t:peaklong}. The ``peak'' at $\ell \simeq +$17\deg (red diamond) is the innermost bin and therefore may not be a genuine local maximum but rather a reflection of the established increase of SFF at small Galactic radii (\citealp{Ragan2016}  and Figure\ \ref{f:SFFvsGLAT}). After removing the AMCs, the overall mean SFF drops to 0.233 (black dashed line in Figure \ref{f:SFF_glong}), and the peaks associated with the Scutum arm ($\ell \sim 31$\deg), 3\,kpc-North arm ($\ell \simeq 25$\deg) and the interarm peak at $\ell \simeq -27$\deg\, are no longer statistically significant. The peaks at $\ell \simeq -22$\deg\, (Perseus arm tangent), $\ell \simeq -32$\deg\, (Norma arm tangent), and the ``peak'' at $\ell \simeq +17$\deg\, all remain statistically significant. Due to the slight drop in mean SFF after the removal of the AMCs, one new peak (i.e. now significant relative to reduced mean SFF) appears at $\ell \simeq -48$\deg, arguably associated with the Centaurus arm tangent. 

As we showed in the previous section, $L_\mathrm{bol}$/$M_\mathrm{tot}$ is related to the SFF but is more strongly affected by the presence of the AMCs. Similar to the SFF, the peaks at $\ell \simeq 31$\deg\, and $-27$\deg\, are completely dominated by AMCs. On the other hand, at greater |$\ell$|, the behaviours of the SFF and $L_\mathrm{bol}$/$M_\mathrm{tot}$ distributions differ. For example, we see elevated $L_\mathrm{bol}$/$M_\mathrm{tot}$ at $\ell \sim -60$\deg, but the corresponding bins in the SFF distribution are below the mean. This means that a small number of sources are dominating the luminosity, which has a weaker effect on SFF. Interestingly, the Sagittarius arm tangent at $\ell \simeq 50^{\circ}$ does not feature in any of these distributions \citep[cf.][]{Benjamin2008,Urquhart2014a}, but all other tangent longitudes exhibit some elevation in SFF above the mean, albeit marginal, owing (mainly, but not exclusively) to the AMCs. 

%%%%%%%%%%%%%%%%%%%%%%%%%%%%%%%%%
\begin{figure*}
\includegraphics[width=\textwidth]{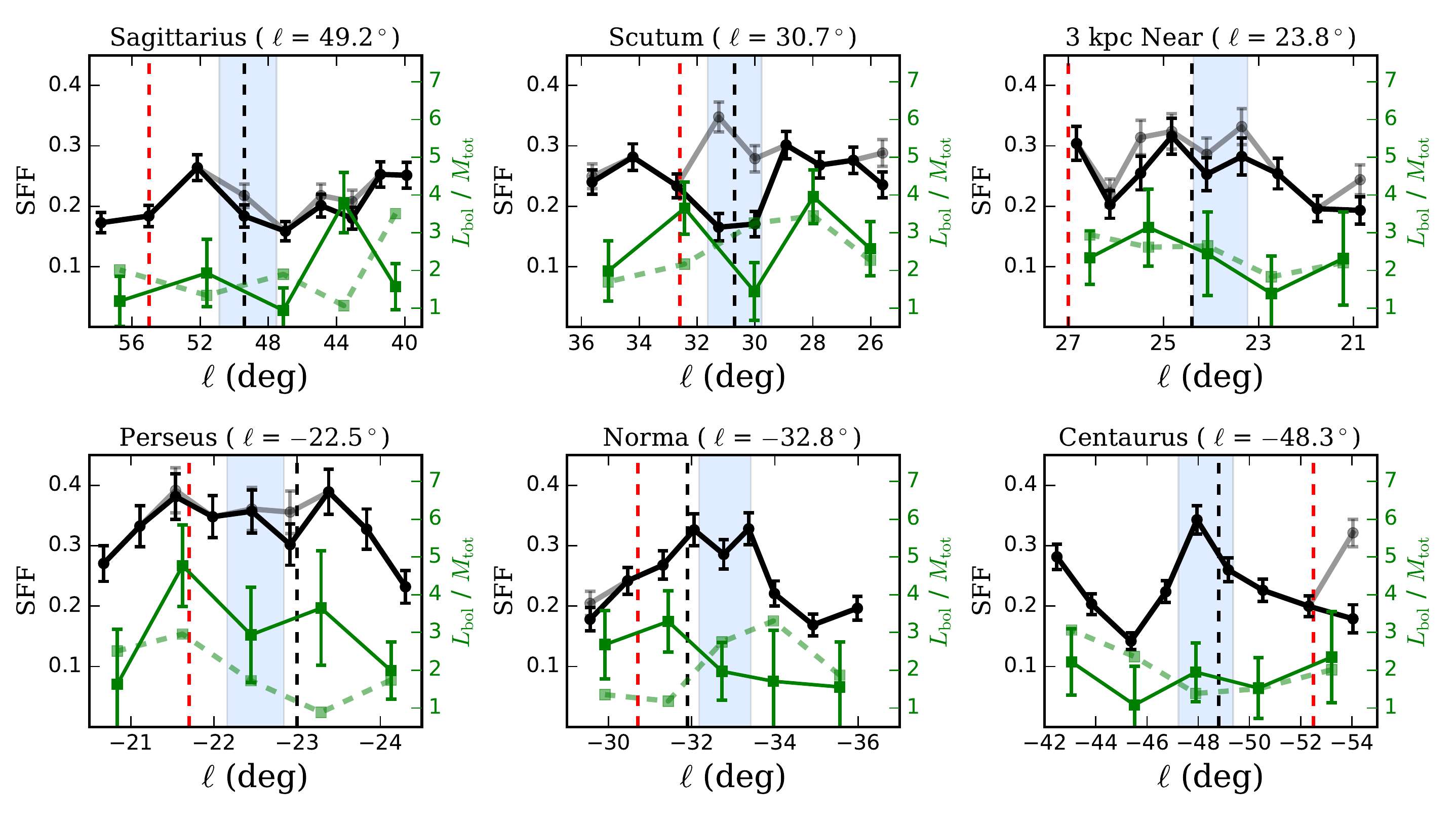}
\caption{Zoom in to the distribution of SFF with longitude near the adopted tangent point longitude locations. The grey line shows the full sample, and the dark black line indicates the SFF after the removal of AMCs. The bins are calculated such that an equal number of sources are in each over the 8 intervals considered at each tangent point. As in Figure \ref{f:SFFvsGLAT}, the green curves represent the $L_\mathrm{bol}$/$M_\mathrm{tot}$ ratio (in units of $L_{\odot}$ / $M_{\odot}$) before and after AMC removal (dashed and solid lines, respectively). The relevant axis values are at the right edge of each panel, and the error bars show the standard error of the mean. The blue shaded region is the peak longitude of the ATLASGAL clump distribution, which we adopt for tangent reference longitude (see Table \ref{t:tangentlong}). The vertical red dashed lines represent the median longitude for stars and the black dashed lines show the median longitude for gas tracers \citep{HouHan2015}. 
\label{f:tangents}}
\end{figure*}

\input{tangent_slopes.tex}

\subsection{Ordering of clump evolutionary stages}

Nearby spiral galaxies exhibit an offset between dust lanes and the evolved stellar population within spiral arms \citep[e.g.][]{Elmegreen1980}. Complementary observations of additional gas and dust tracers (e.g. \hone, radio continuum) hint at a complex segregation of evolutionary stages within arms \citep[e.g.][]{Kaufman1989}, but it is difficult to assess this in detail in external galaxies due to resolution limitations. Nevertheless, these offsets are thought to be the consequence of the way in which gas in spiral galaxies cycles through the disc with respect to the potential of the spiral arms. 
Material enters an arm from the interior side (i.e., the edge nearest to the Galactic centre), molecular clouds and stars form, and material leaves the arm from the exterior side. The observational expectation from this is that the early evolutionary stages (i.e., gas and dust) should be interior to the later stages (i.e., stars). In the Milky Way, such segregation between early and late stages would be most evident at the arm tangents, where adjacent features within an arm would translate to offsets in longitude. Offsets between the peaks in tracers of gas and stars of 1.3\deg\, - 5.8\deg\, are observed at the spiral arm tangent points of the Milky Way \citep{HouHan2015}. 

In our study, we aim to investigate whether an offset in evolutionary stages is imprinted at the clump stage. Since the time scales associated with the pre-stellar and proto-stellar clump phases ($\ls$ a few times $10^5$ years) are shorter than the expected arm-crossing time ($\gs$ $10^6$ years), if there is significant triggering of star formation by the spiral arm potential, we expect to see consistent variations in SFF across the arm tangents.

We show zoom-ins of the SFF distribution near the tangent-point longitudes in Figure \ref{f:tangents}. The black and red dashed vertical lines represent the peak longitudes of the gas and stellar population, respectively, found by \citet{HouHan2015}. The Sagittarius, Scutum, 3\,kpc-Near and Centaurus tangent points exhibit the expected sense of the offset between the gas and stellar population, with the stellar population peak exterior (i.e. at longitudes further from the Galactic Centre) to the peak in gas tracers. The Perseus and Norma arms show the reverse sense of the expected offset (stellar peak at larger |$\ell$| compared to the gas peak), though the meaning of this reversal was not explained in \citet{HouHan2015}.

The Galactocentric radius-dependent width \citep{Reid2014a} of each arm is again represented by the blue shaded regions, centred on the longitude of the ATLASGAL source peak \citep{Beuther2012, HouHan2015}. Each section has nine longitude bins spaced such that they contain equal numbers of clumps, distributed in the range $\pm$3 times the arm width.  As in Figure \ref{f:SFF_glong}, the grey curves show the full catalogue of Hi-GAL clumps, and the thick black line shows the distribution after the AMCs are removed. Of the six plots, only that centred on the Centaurus tangent shows any hint of an enhancement in SFF, with respect to the inter-arm sources, that falls within the arm.  Also, we see again that the removal of the AMC sources diminishes or removes altogether any minor SFF peaks associated with the other tangent points.

If compact Hi-GAL sources were to show spatial segregation with their evolutionary stage through an arm, this should appear as a gradient in SFF versus longitude at the tangent points. The na\"{i}ve expectation would be that the SFF value would have a positive gradient on some scale between the gas peak and the stellar peak (i.e. from the black dashed line toward the red). To determine whether there are any statistically significant trends in our data, we perform the Spearman rank test on the relation between SFF and longitude over varying longitude ranges: 3 (shown in Figure \ref{f:tangents}), 2.5 and 2 times the width of the arm (see Table \ref{t:tangentlong}). In all cases, the $p$ values for the SFF versus longitude relations across the spiral arm tangents both before and after the removal of the AMCs, summarised in Table \ref{t:tangentslopes}, are well above the acceptable threshold for a correlation, regardless of the longitude range used. We therefore see no evidence of segregation in evolutionary phase at the clump stage, which is either because it is absent or it is unresolved.

%%%%%%%%%%%%%%%%%%%%%%%%%%%%%%%%%
%%%%%%%%%%%%%%%%%%%%%%%%%%%%%%%%%
\section{Discussion}

\subsection{Global trends in Milky Way star formation}

We have studied how the number and average star-forming properties of Hi-GAL compact sources vary as a function of position in the inner Galaxy and at the spiral-arm tangent-point longitudes. We have also identified the subset of Hi-GAL compact sources associated with the known 22 most massive ATLASGAL clusters (AMCs; \citealp{Urquhart2017}) in the inner Galaxy (see Table \ref{t:atlasgal}), most of which are located at Galactic longitudes $\ell < 40$\deg. The AMCs account for only 7.6\% of the Hi-GAL compact clump catalogue sources. Since the AMCs account for a large fraction of the mass and luminosity in the inner Galaxy \citep{Urquhart2017}, and could bias the spatially-averaged evolutionary metrics, we examine the effect of excluding them from the averages.

There are statistically significant but small ($< 50$\%) localised increases in the fraction of Hi-GAL sources that are infrared-bright and therefore show evidence of active star formation associated with the locations of the tangents of the inner spiral arms. This fraction \citep[the SFF;][]{Ragan2016} is related to the mean evolutionary state of a given region but is potentially significantly affected by the presence of very high-mass and high-luminosity massive young stellar objects (MYSOs), i.e., the pre-cursors of clusters containing O-type stars, since these have very short pre-stellar lifetimes, becoming IR-bright in as little as a few tens of thousands of years (\citealp{Mottram2011}, \citealp{Urquhart2017}). This is illustrated in Table \ref{t:atlasgal}, which lists SFF values for the AMCs. After removing the AMCs from the sample, Figure \ref{f:SFF_glong} shows that some of the peaks are diminished, but some inner-arm tangents, generally the innermost, still have mean SFF values more than 3\,$\sigma$ above the sample mean after removal of the AMCs.

Figure \ref{f:SFFvsGLAT} shows the distribution of SFF as a function of Galactic latitude. The distribution of the whole sample (Fig.\ \ref{f:SFFvsGLAT}a) peaks close to $b$\,=\,$0$ and this is consistent with the fact that the faster-evolving higher-mass MYSOs (and OB stars) have a small scale height above the Plane ($\sim$20-30\,pc in the inner Galaxy: \citealp{Reed2000,Urquhart2013}, which is 0.2-0.3 degrees at a median distance of $\sim$6\,kpc).  The pattern inside $|\, \ell \,| < 40^{\circ}$ (Fig. \ref{f:SFFvsGLAT}b) is similar but outside this longitude range the distribution of SFF is much flatter or truncated within $|\,b\,| \le 0.5^{\circ}$ (Fig. \ref{f:SFFvsGLAT}c).  The average also falls in this outer region (to $\sim$0.23), indicating that the flatter distribution is not simply due to an increased scale height. This suggests that the fraction of sources containing the most massive YSOs is lower outside the innermost regions of the Galaxy and, specifically, the region associated with the inner spiral arms and the area swept by the Galactic bar. This further implies that the young clusters forming in the outer Galaxy and above the plane are smaller and are lacking in the highest-mass and highest-luminosity stars. This would be consistent with the results of \citet{Pflamm-Altenburg2013}, who examined the cluster mass function in M33.  Additionally, if most stars form in clusters, then the average initial mass function (IMF) of stars is the result of the convolution of the cluster mass function with the IMF forming within clusters. \citet{Pflamm-Altenburg2007} suggested that the highest-mass star that forms in a cluster depends on the cluster mass. Since this would result in truncated IMFs within smaller clusters, it would imply that the average IMF in the Galaxy must be somewhat position-dependent. This is a potential explanation for the declining SFF gradient found in \citet{Ragan2016}, i.e., that there is a gradient in the locally averaged IMF of currently forming stars with Galactocentric radius. However, studies of the H$\alpha$ to FUV ratio in samples of nearby galaxies (e.g., \citealp{Fumigalli2011}, \citealp{Hermanowicz2013}) and UV surface brightness in M83 \citep{Koda2012} have found results more consistent with an IMF that is fully stochastically sampled everywhere and not truncated by cluster mass.

As mentioned above, the value of the SFF may also depend on the (mean) evolutionary state of a sample in that a large amount of incipient star formation in the form of pre-stellar clumps without infrared emission would reduce it, and vice versa.  However, it is unlikely that there are evolutionary phase variations and gradients that are coherent on such large scales and significant large-scale variations in related star-formation metrics have not been found (\citealp{Moore2012}, \citealp{Eden2012, Eden2015}, etc.).  It therefore seems likely that SFF in large, comparable-sized samples principally traces the relative population of massive YSOs. 

Although the extreme AMC sources are predominantly found in the inner Plane, the large-scale variations in SFF in Figures \ref{f:SFFvsGLAT} and \ref{f:SFF_glong} are not due to their presence, as their removal has little effect on the SFF distributions, beyond the reduction in significance or removal of one or two peaks in the longitude plot (Figure \ref{f:SFF_glong}), including that potentially associated with the Scutum arm tangent. This is because the sample of AMC-related Hi-GAL sources are a small fraction (7.6\%) of the larger sample. 

\subsection{Star formation in spiral arms and the bar}

We have shown that some tangent-point longitudes exhibit moderate coincident enhancement in SFF while some do not and that several peaks can be explained by the presence of AMCs (see Figure \ref{f:SFF_glong}), even though the AMCs only account for a small fraction of the total number of sources. Even the peaks in SFF that persist after removing the AMCs are only statistically significant at the $\sim$3-$\sigma$ level, which amounts to only a 20-50\% elevation above the mean. 

The most massive sites of star formation in the Galaxy (the AMCs) are indeed largely found at the tangent-point longitudes, which themselves have been defined by the peaks in longitude distributions of various tracers. This poses a bit of a dilemma: are most of the AMCs found at tangent longitudes, or have we defined the tangent directions to be toward the AMCs? We have attempted to circumvent the dominating effects of the AMCs by examining the averages without them, and we find that the tangents are weaker features in the longitude distribution (as expected) but the overall negative gradient in SFF with increasing |$\ell$| remains, which is consistent with the radial gradient characterised in \citet{Ragan2016}.

It is also worth considering that all but one of the peaks in Figure \ref{f:SFF_glong} with $\gs 3$-$\sigma$ increases in SFF are located at longitudes $|\,\ell\,| \ls 35^{\circ}$, inside the $\sim$4\,kpc radius swept by the Galactic bar. This makes them likely to be associated with the inner spiral features expected to be driven by the rotating bar potential (e.g., \citealp{Sormani2015}, \citealp{Li2016}). Such features appear in these (non-self-gravitating) hydrodynamic models as kinematic density waves formed due to small oscillations around the otherwise closed, elliptical, ballistic $x_1$ orbits. Furthermore, the fiducial model of \cite{Sormani2015} shows strong arm-tangent features at $\ell \sim 15^{\circ}$ and $\sim -22^{\circ}$, close to the two observed SFF enhancements that survive the removal of extreme sources from the sample (the starred features in Figure \ref{f:SFF_glong}).  A scenario in which the Galactic bar is the dominant player affecting trends in SFF in the Galactic plane is not inconsistent with the findings of Paper I, if the bar's influence diminishes with increasing Galactocentric radius. 

In any case, the very modest and somewhat inconsistent enhancements in SFF associated with arm tangents suggests that the spiral arms in the Milky Way do not have a strong effect on the star formation occurring within dense clumps. It may be that there is a larger influence on the formation and internal structure of molecular clouds that can be attributed to spiral-arm density-wave shocks or the fall into the arm potential and that, once dense clumps have formed within clouds, they are disconnected from their environment.  That possibility is a topic for further study.

\subsection{No evidence for ordering of clump evolutionary stage}

The offsets observed between gas and stars in the arms of spiral galaxies are often interpreted as a consequence of the movement of gas relative to the spiral potential, wherein gas flows into the spiral potential from the interior edge of the arm, molecular clouds and stars form -- possibly triggered by shock compression induced by a spiral density wave \citep{Roberts1969} -- and the downstream lane of stars represents the end product of this process. For the Milky Way, \citet{HouHan2015} have shown that (except in the cases of the Perseus and Norma tangencies) the stellar population peaks at larger |\,$\ell$\,|, in agreement with expectation. We have examined the mean clump evolutionary stage (as measured by the SFF) at the tangent longitudes to see whether there is a corresponding gradient in a similar sense. We find no evidence for any such trend across the spiral arms. 

The absence of any significant gradient in SFF with longitude across the arms could have a number of explanations. It could mean that the timescale over which a given Hi-GAL clump becomes actively star-forming (and thus infrared-bright) after being shock-compressed is too short over the scales we are probing (0.4 - 0.9\,kpc) for there to be a measurable pattern from one side of the arm to the other, in other words the variations may be unresolved.  It could also mean that star formation is more rapid than the arm-crossing time but remains stochastic across the arm, i.e., unaffected by the arm passage. Indeed, even the AMCs do not show a preference to be on one side or the other of the tangent point with respect to the stellar peak / downstream side of the arm (see Figure \ref{f:tangents}). Additionally, galaxy simulations have shown that the dynamics of large-scale molecular cloud complexes, upon entering the spiral arm potential well after inter-arm passage, are dynamically disrupted and fragmented, and they do not experience a marked enhancement in star formation \citep{DuarteCabral2017}.
We also note that the presence of a bar in the inner 3-4\,kpc of the Galaxy could have a complicating effect on whether any clear sequencing of evolutionary stages would persist long enough for an observable effect. \citet{Dobbs2010} find that, in a simulated barred galaxy, the stellar clusters spanning a range of ages of $\sim$50\,Myr are completely spatially mixed, indicating that when a bar is present, it dominates the dynamics much more than regular spiral galactic potentials.

%%%%%%%%%%%%%%%%%%%%%%%%%%%%%%%%%
%%%%%%%%%%%%%%%%%%%%%%%%%%%%%%%%%
\section{Summary and Conclusions}

The relationship between spiral structure and star formation has long been a debate based largely on evidence drawn from observations of nearby galaxies. With a growing number of unbiased surveys of the Milky Way plane, we are now equipped to investigate this question within our own Galaxy. The Hi-GAL survey in particular has given us a new perspective on the distribution of clumps at the earliest phase of star formation throughout the Galaxy, providing a comprehensive catalogue of compact, star-forming clumps numbering over 10$^5$ in the inner Galaxy alone. We address long-standing questions that have, in the past, been plagued by small number statistics: do the Milky Way's spiral arms play a role in star formation?

We use the new Hi-GAL compact source catalogue for the inner Galaxy \citep{Elia2017} to study how the properties of compact clumps vary as a function of position with respect to spiral arms. We use the star-forming fraction \citep[SFF;][]{Ragan2016} as a useful measure of the prevalence of star formation in a given area. The SFF positively correlates with commonly-used metrics of clump evolutionary stage such as the luminosity to mass ratio and the ratio of total to sub-millimetre luminosities. Compared to these clump-scale metrics, the SFF has the advantage that its value will not be dominated by a small number of very luminous sources, but rather lets us look at the broad evolutionary trends across the mass spectrum of dense clumps such that we can measure the spatial variations of clump properties more widely. 

Using recent results from the ATLASGAL survey in which \citet{Urquhart2017} identified the most massive clusters in the inner Galactic plane (referred to in this work as the ATLASGAL massive clusters, or AMCs), we identify the Hi-GAL clump counterparts of the AMCs in order to examine trends in the positional distributions with and without their dominating presence. The modest enhancements in SFF at tangent point longitudes present when the AMCs are included largely disappear with their removal, leaving a flatter distribution of SFF with longitude, indicating that the prevalence of star formation in most Hi-GAL clumps (at least in the 92.4\% of the catalogue that do not reside in AMCs) does not change appreciably with position relative to a spiral arm. 

We also find no gradient in the SFF with position across spiral arms. If the observed offset between stars and gas in the Milky Way \citep{HouHan2015} and nearby galaxies is a consequence of the time ordering of star formation predicted by density wave theory, then it is not evident within the clump stage. It may be the case that the clump phase does not encompass the full relevant time scale over which cycle of transforming gas into stars takes place, or it may be because the star formation happening in Hi-GAL clumps is stochastic and not coherently ordered by a density wave.

\section*{Acknowledgements} 
This research has made use of NASA Astrophysics Data System and Astropy, a community-developed core Python package for Astronomy \citep{astropy}.  SER acknowledges support from the European Union's Horizon 2020 research and innovation programme under the Marie Sk{\l}odowska-Curie grant agreement \# 706390 and the {\sc vialactea} Project, a Collaborative Project under Framework Programme 7 of the European Union, funded under Contract \# 607380.

\bibliographystyle{mnras}
\bibliography{tangents}

\appendix
\section{Maps of evolutionary tracers}

\begin{figure*}
\centering
\includegraphics[width=\textwidth]{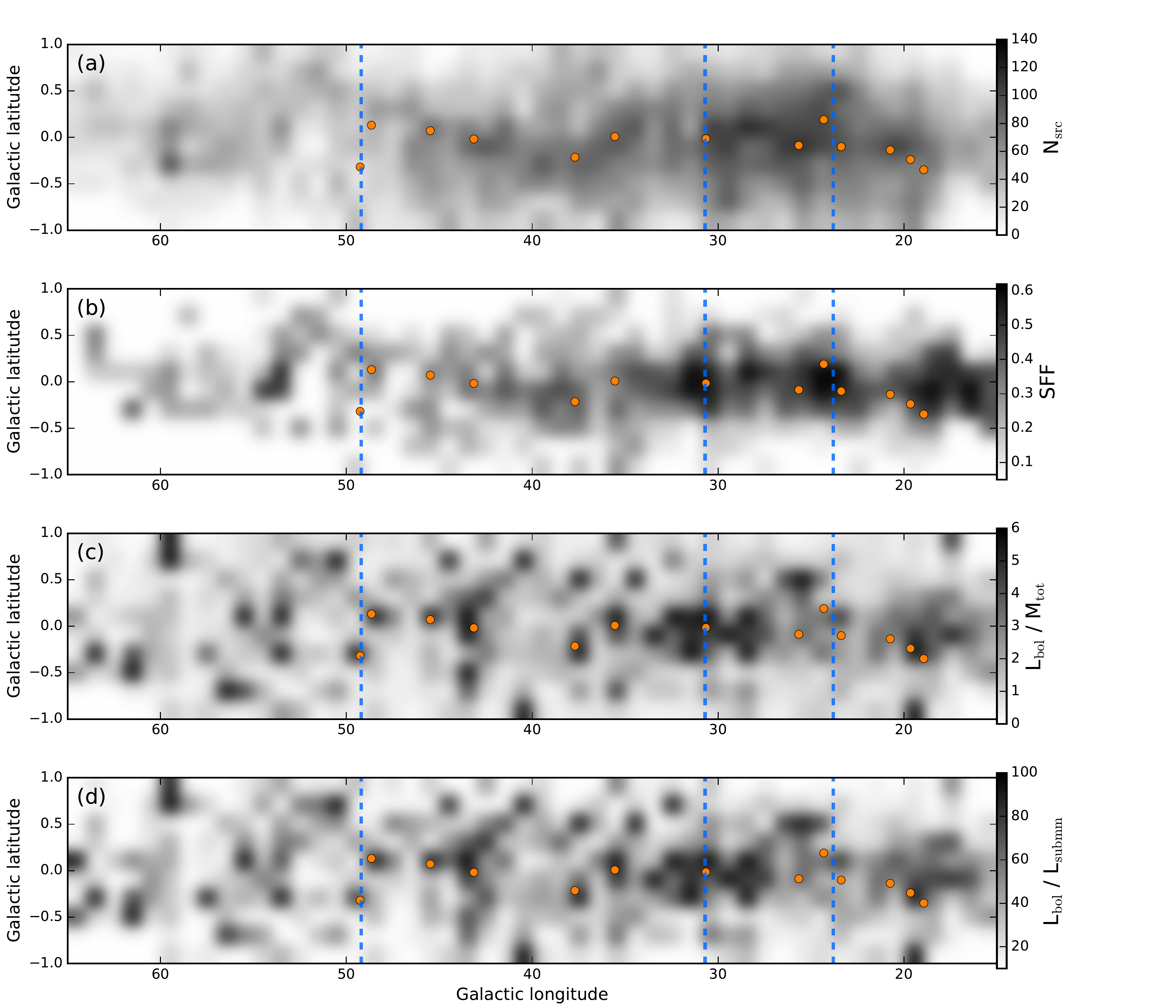} 
\caption{(a) Two-dimensional histogram of the Hi-GAL source counts in the 15\deg $< \ell <$ 65\deg portion of the first quadrant of the Milky Way, excluding the CMZ. Bins are 0.2\deg wide in latitude and 1.0\deg wide in longitude.  The orange circles are the locations of 22 most luminous ATLASGAL sources (see Table \ref{t:atlasgal}). Vertical dashed blue lines indicate the longitudes of spiral arm tangent points (see Table~\ref{t:tangentlong}). (b) Two-dimensional histogram of SFF. (c): Two-dimensional histogram of luminosity-to-mass ratio (in units of $L_{\odot}$ / $M_{\odot}$).  (d): Two-dimensional histogram of the $L_\mathrm{bol}$ / $L_\mathrm{submm}$ ratio.  \label{f:SFFmap1}}
\end{figure*}

\begin{figure*}
\centering
\includegraphics[width=\textwidth]{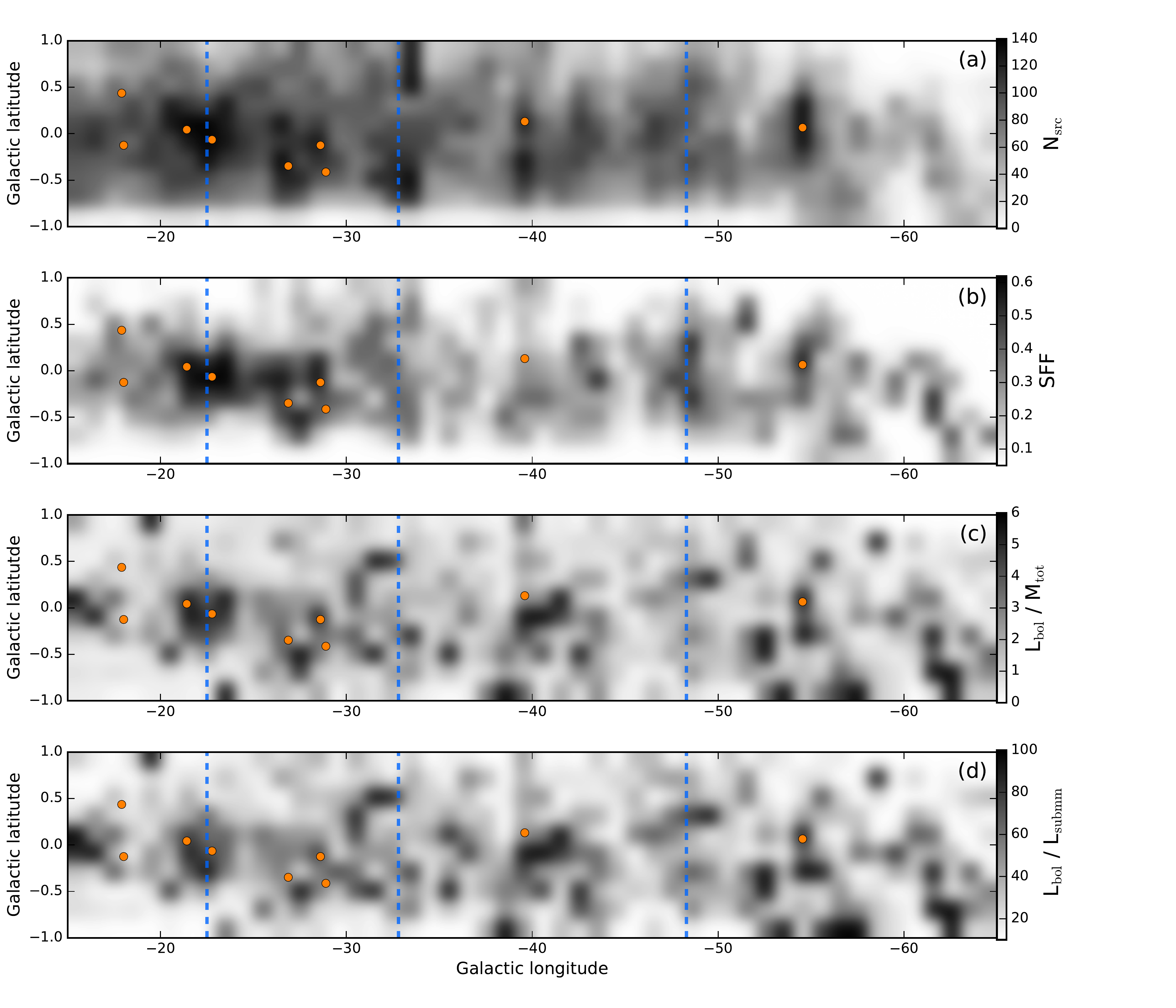}
\caption{Same as Figure \ref{f:SFFmap1} but for the $-$65\deg $< \ell <$ $-$15\deg portion of the fourth quadrant of the Milky Way.  \label{f:SFFmap4}}
\end{figure*}

We show in Figures \ref{f:SFFmap1} and \ref{f:SFFmap4} the distribution of (a) the number of Hi-GAL clumps, (b) the SFF, (c) the $L_\mathrm{bol}$ / $M_\mathrm{tot}$ ratio and (d) the $L_\mathrm{bol}$ / $L_\mathrm{submm}$ ratio of the first and fourth Galactic quadrants, respectively. The values are computed over equal areas ($\ell \times b$ = 1\deg\, $\times$\, 0.2\deg). We show the adopted locations of the spiral arm tangent points (vertical dashed blue lines) and the locations of the centres of the AMCs (see Appendix \ref{a:atlasgal}) in the orange circles.

\section{Most massive ATLASGAL clusters}
\label{a:atlasgal}

Using the unbiased ATLASGAL survey, we use the \citet{Urquhart2014c} catalogue of compact sources to identify the brightest star-forming complexes in the Galaxy. Within the longitude range considered in this paper (15\deg\, $<\,|\,\ell\,|\,<$ 65\deg), there are 25 exceeding X\,$L_{\odot}$, the boundaries of which are listed in Table \ref{t:atlasgal} along with the number of Hi-GAL compact clumps and overall SFF in the region. These regions represent the most active star formation in the Galaxy and encompass 7.6\% of the total Hi-GAL compact clump catalogue. We refer to these sources throughout the text as the ATLASGAL Massive Clusters (AMCs). We study the properties of the Hi-GAL clump catalogue with and without the AMCs in order to ascertain their importance to the Galaxy's star formation budget.

\input{atlasgal_removed.tex}

% Don't change these lines
\bsp	% typesetting comment
\label{lastpage}
\end{document}

%% file: houhan_tangents_props_tab.tex
\begin{table}
\caption{Tangent longitudes and dimensions.}
\begin{tabular}{lrccc}
\hline 
Arm & $\ell^\mathrm{a}~~$   & $D^\mathrm{b}$ & Arm Width$^\mathrm{c}$ & Ang. Width \\
     & (deg)           &  (kpc)         & (pc)                   &  (deg) \\
\hline 
Sagittarius    &  49.2 & 4 & 234 & 3.4 \\
Scutum         &  30.7 & 5 & 163 & 1.9 \\
3\,kpc (near)  &  23.8 & 6 & 119 & 1.1 \\
Perseus      & $-$22.5 & 8 &  95 & 0.7 \\
Norma        & $-$32.8 & 7 & 151 & 1.2 \\
Centaurus    & $-$48.3 & 6 & 224 & 2.1 \\
\hline
\end{tabular}
\\
$^\mathrm{a}$ Taken from \citet{HouHan2015}, based on ATLASGAL sources.\\
$^\mathrm{b}$ Approximate distance to tangent point from \citet{Vallee2014c}.\\
$^\mathrm{c}$ Based on the \citet{Reid2014a} model.
\label{t:tangentlong}
\end{table}

%% file: peak_longitudes.tex
\begin{table}
\caption{Peak SFF longitudes}
\begin{tabular}{cccl}
$\ell$ (\deg) & SFF & L/M & Associated Arm \\
\hline
$-$48$^\mathrm{a}$ & 0.29 & 1.8 & Centaurus \\
$-$32$^\mathrm{b}$ & 0.30 & 2.8 & Norma \\
$-$27 & 0.32 & 3.5 & \\
$-$22$^\mathrm{b}$ & 0.35 & 3.7 & Perseus \\\
+17$^\mathrm{c}$ & 0.38 & 2.6 & \\
+25 & 0.31 & 2.9 & 3kpc (near) \\
+31 & 0.32 & 4.5 & Scutum \\
\hline
\end{tabular}
\\
$^\mathrm{a}$ Peak appears after removal of AMCs. \\
$^\mathrm{b}$ Peak remains after removal of AMCs. \\
$^\mathrm{c}$ Innermost bin.
\label{t:peaklong}
\end{table}

%% file: tangent_slopes.tex
\begin{table}
\caption{Evolutionary trends across tangent longitudes.}
\begin{tabular}{lcccc}
\hline
Arm & Ang. range$^\mathrm{a}$ & N$_\mathrm{src}$  & $p_\mathrm{SFF}^\mathrm{b}$ & $p_\mathrm{SFF}^\mathrm{b,c}$ \\
& (deg) & (bin$^{-1}$) & & \\
\hline 
\multicolumn{2}{l}{$\pm$ 3 times arm width} & & \\
\hline
Sagittarius  & 17.9 &  713  & 0.26  & 0.41 \\
Scutum       & 10.1 &  754  & 0.46  & 0.77  \\
3\,kpc (near) & 6.0 &  501  & 0.38  & 0.10 \\
Perseus      &  3.6 &  389  & 0.73  & 0.73 \\
Norma        &  6.4 &  617  & 0.41  & 0.64 \\
Centaurus    & 11.6 &  832  & 0.64  & 0.52  \\
\hline
\multicolumn{2}{l}{$\pm$ 2.5 times arm width} & & \\
\hline
Sagittarius  & 14.9 &  578  & 0.73  & 0.70 \\
Scutum       &  8.3 &  611  & 0.90  & 0.83  \\
3\,kpc (near) & 5.0 &  422  & 0.46  & 0.52 \\
Perseus      &  3.0 &  324  & 0.83  & 0.86 \\
Norma        &  5.3 &  527  & 0.55  & 0.55 \\
Centaurus    &  9.3 &  661  & 0.85  & 0.85  \\
\hline
\multicolumn{2}{l}{$\pm$ 2 times arm width} & & \\
\hline
Sagittarius  & 10.3 &  451  & 0.86  & 1.0 \\
Scutum      &   6.5 &  478  & 0.59  & 0.36  \\
3\,kpc (near) & 4.0 &  344  & 0.41  & 1.0 \\
Perseus      &  2.4 &  258  & 0.85  & 0.80 \\
Norma       &  4.2  &  444  & 0.33  & 0.33 \\
Centaurus   &  7.2  &  542  & 0.22  & 0.22  \\
\hline
\end{tabular}
\\
$^\mathrm{a}$ Total angular scale over which Spearman rank was calculated, spanning indicated range in multiples of the angle subtended by the width of the arm (see Table \ref{t:tangentlong}).\\
$^\mathrm{b}$ $p$ value of the Spearman rank test.\\
$^\mathrm{c}$ Values after AMCs are removed (see Table \ref{t:atlasgal}).
\label{t:tangentslopes}
\end{table}

%% file: atlasgal_removed.tex
\begin{table*}
\caption{Most massive ATLASGAL clusters (AMCs) identified in \citet{Urquhart2017}.}
\begin{center}
\begin{tabular}{lccrrrl}
Cluster & Min. GLON & Max. GLON & Min. GLAT & Max. GLAT & N$_\mathrm{HIGAL}^\mathrm{a}$ & SFF \\
Name & (deg) & (deg) & (deg) & (deg) & & \\
\hline
G305.453+00.065  &   $-$54.911  &  $-$53.789  &  $-$0.357  &  +0.504  &  559  &  0.45  \\
G320.403+00.131  &   $-$39.633  &  $-$39.566  &  +0.089  &  +0.177  &  10  &  0.80  \\
G331.104$-$00.413  &   $-$29.326  &  $-$28.541  &  $-$0.541  &  $-$0.302  &  105  &  0.49  \\
G331.394$-$00.125  &   $-$29.071  &  $-$28.039  &  $-$0.382  &  +0.262  &  382  &  0.52  \\
G333.125$-$00.348  &   $-$28.038  &  $-$26.037  &  $-$0.854  &  +0.294  &  1243  &  0.43  \\
G337.228$-$00.065  &   $-$22.887  &  $-$22.651  &  $-$0.174  &  +0.077  &  57  &  0.70  \\
G338.586+00.043  &   $-$21.458  &  $-$21.371  &  $-$0.016  &  +0.142  &  13  &  0.69  \\
G341.982$-$00.125  &   $-$18.068  &  $-$17.936  &  $-$0.174  &  $-$0.057  &  6  &  0.50  \\
G342.089+00.436  &   $-$18.036  &  $-$17.727  &  +0.397  &  +0.512  &  29  &  0.48  \\
G018.929$-$00.343  &   +18.659  &  +19.223  &  $-$0.777  &  +0.129  &  232  &  0.43  \\
G019.649$-$00.239  &   +19.609  &  +19.706  &  $-$0.266  &  $-$0.172  &  6  &  0.67  \\
G020.739$-$00.136  &   +20.654  &  +20.892  &  $-$0.359  &  $-$0.012  &  55  &  0.65  \\
G023.375$-$00.101  &   +23.086  &  +23.582  &  $-$0.337  &  +0.114  &  107  &  0.51  \\
G024.319+00.191  &   +24.002  &  +24.561  &  +0.014  &  +0.351  &  100  &  0.47  \\
G025.656$-$00.087  &   +25.328  &  +25.983  &  $-$0.189  &  +0.106  &  116  &  0.58  \\
G030.650$-$00.015  &   +29.779  &  +31.784  &  $-$0.417  &  +0.454  &  674  &  0.49  \\
G035.553+00.008  &   +35.476  &  +35.602  &  $-$0.094  &  +0.116  &  16  &  0.69  \\
G037.702$-$00.214  &   +37.341  &  +38.039  &  $-$0.399  &  $-$0.053  &  96  &  0.52  \\
G043.141$-$00.018  &   +43.061  &  +43.236  &  $-$0.077  &  +0.044  &  35  &  0.74  \\
G045.486+00.071  &   +45.421  &  +45.549  &  $-$0.032  &  +0.141  &  26  &  0.65  \\
G048.651+00.131  &   +48.579  &  +48.851  &  +0.007  &  +0.246  &  25  &  0.56  \\
G049.261$-$00.318  &   +48.843  &  +49.669  &  $-$0.516  &  +0.028  &  56  &  0.45  \\
\hline
\end{tabular}
\end{center}
$^\mathrm{a}$\, Number of Hi-GAL sources in the area enclosed in the area bounded by Min./Max. GLON and Min./Max. GLAT. \label{t:atlasgal}
\end{table*}

%% file: ragan_tangents.bbl
\begin{thebibliography}{}
\makeatletter
\relax
\def\mn@urlcharsother{\let\do\@makeother \do\$\do\&\do\#\do\^\do\_\do\%\do\~}
\def\mn@doi{\begingroup\mn@urlcharsother \@ifnextchar [ {\mn@doi@}
  {\mn@doi@[]}}
\def\mn@doi@[#1]#2{\def\@tempa{#1}\ifx\@tempa\@empty \href
  {http://dx.doi.org/#2} {doi:#2}\else \href {http://dx.doi.org/#2} {#1}\fi
  \endgroup}
\def\mn@eprint#1#2{\mn@eprint@#1:#2::\@nil}
\def\mn@eprint@arXiv#1{\href {http://arxiv.org/abs/#1} {{\tt arXiv:#1}}}
\def\mn@eprint@dblp#1{\href {http://dblp.uni-trier.de/rec/bibtex/#1.xml}
  {dblp:#1}}
\def\mn@eprint@#1:#2:#3:#4\@nil{\def\@tempa {#1}\def\@tempb {#2}\def\@tempc
  {#3}\ifx \@tempc \@empty \let \@tempc \@tempb \let \@tempb \@tempa \fi \ifx
  \@tempb \@empty \def\@tempb {arXiv}\fi \@ifundefined
  {mn@eprint@\@tempb}{\@tempb:\@tempc}{\expandafter \expandafter \csname
  mn@eprint@\@tempb\endcsname \expandafter{\@tempc}}}

\bibitem[\protect\citeauthoryear{{Andr{\'e}}, {Ward-Thompson}  \&
  {Barsony}}{{Andr{\'e}} et~al.}{1993}]{Andre1993}
{Andr{\'e}} P.,  {Ward-Thompson} D.,   {Barsony} M.,  1993, \mn@doi [\apj]
  {10.1086/172425}, \href {http://adsabs.harvard.edu/abs/1993ApJ...406..122A}
  {406, 122}

\bibitem[\protect\citeauthoryear{{Astropy Collaboration} et~al.,}{{Astropy
  Collaboration} et~al.}{2013}]{astropy}
{Astropy Collaboration} et~al., 2013, \mn@doi [\aap]
  {10.1051/0004-6361/201322068}, \href
  {http://adsabs.harvard.edu/abs/2013A%26A...558A..33A} {558, A33}

\bibitem[\protect\citeauthoryear{{Battisti} \& {Heyer}}{{Battisti} \&
  {Heyer}}{2014}]{Battisti2014}
{Battisti} A.~J.,  {Heyer} M.~H.,  2014, \mn@doi [\apj]
  {10.1088/0004-637X/780/2/173}, \href
  {http://adsabs.harvard.edu/abs/2014ApJ...780..173B} {780, 173}

\bibitem[\protect\citeauthoryear{{Benjamin}}{{Benjamin}}{2008}]{Benjamin2008}
{Benjamin} R.~A.,  2008, in {Beuther} H.,  {Linz} H.,   {Henning} T.,  eds,
  Astronomical Society of the Pacific Conference Series Vol. 387, Massive Star
  Formation: Observations Confront Theory. p.~375

\bibitem[\protect\citeauthoryear{{Benjamin} et~al.,}{{Benjamin}
  et~al.}{2005}]{Benjamin2005}
{Benjamin} R.~A.,  et~al., 2005, \mn@doi [\apjl] {10.1086/491785}, \href
  {http://adsabs.harvard.edu/abs/2005ApJ...630L.149B} {630, L149}

\bibitem[\protect\citeauthoryear{{Beuther} et~al.,}{{Beuther}
  et~al.}{2012}]{Beuther2012}
{Beuther} H.,  et~al., 2012, \mn@doi [\apj] {10.1088/0004-637X/747/1/43}, \href
  {http://adsabs.harvard.edu/abs/2012ApJ...747...43B} {747, 43}

\bibitem[\protect\citeauthoryear{{Binney} \& {Tremaine}}{{Binney} \&
  {Tremaine}}{1987}]{BinneyTremaine}
{Binney} J.,  {Tremaine} S.,  1987, {Galactic dynamics}.
Princeton University Press

\bibitem[\protect\citeauthoryear{{Choi}, {Dalcanton}, {Williams}, {Weisz},
  {Skillman}, {Fouesneau}  \& {Dolphin}}{{Choi} et~al.}{2015}]{Choi2015}
{Choi} Y.,  {Dalcanton} J.~J.,  {Williams} B.~F.,  {Weisz} D.~R.,  {Skillman}
  E.~D.,  {Fouesneau} M.,   {Dolphin} A.~E.,  2015, \mn@doi [\apj]
  {10.1088/0004-637X/810/1/9}, \href
  {http://adsabs.harvard.edu/abs/2015ApJ...810....9C} {810, 9}

\bibitem[\protect\citeauthoryear{{Dame} \& {Thaddeus}}{{Dame} \&
  {Thaddeus}}{2008}]{Dame2008}
{Dame} T.~M.,  {Thaddeus} P.,  2008, \mn@doi [\apjl] {10.1086/591669}, \href
  {http://adsabs.harvard.edu/abs/2008ApJ...683L.143D} {683, L143}

\bibitem[\protect\citeauthoryear{{Dobbs} \& {Baba}}{{Dobbs} \&
  {Baba}}{2014}]{DobbsBaba2014}
{Dobbs} C.,  {Baba} J.,  2014, \mn@doi [\pasa] {10.1017/pasa.2014.31}, \href
  {http://adsabs.harvard.edu/abs/2014PASA...31...35D} {31, e035}

\bibitem[\protect\citeauthoryear{{Dobbs} \& {Pringle}}{{Dobbs} \&
  {Pringle}}{2010}]{Dobbs2010}
{Dobbs} C.~L.,  {Pringle} J.~E.,  2010, \mn@doi [\mnras]
  {10.1111/j.1365-2966.2010.17323.x}, \href
  {http://adsabs.harvard.edu/abs/2010MNRAS.409..396D} {409, 396}

\bibitem[\protect\citeauthoryear{{Dobbs}, {Burkert}  \& {Pringle}}{{Dobbs}
  et~al.}{2011}]{Dobbs2011}
{Dobbs} C.~L.,  {Burkert} A.,   {Pringle} J.~E.,  2011, \mn@doi [\mnras]
  {10.1111/j.1365-2966.2011.19346.x}, \href
  {http://adsabs.harvard.edu/abs/2011MNRAS.417.1318D} {417, 1318}

\bibitem[\protect\citeauthoryear{{Duarte-Cabral} \& {Dobbs}}{{Duarte-Cabral} \&
  {Dobbs}}{2017}]{DuarteCabral2017}
{Duarte-Cabral} A.,  {Dobbs} C.~L.,  2017, \mn@doi [\mnras]
  {10.1093/mnras/stx1524}, \href
  {http://adsabs.harvard.edu/abs/2017MNRAS.470.4261D} {470, 4261}

\bibitem[\protect\citeauthoryear{{Dunham}, {Crapsi}, {Evans}, {Bourke},
  {Huard}, {Myers}  \& {Kauffmann}}{{Dunham} et~al.}{2008}]{Dunham2008}
{Dunham} M.~M.,  {Crapsi} A.,  {Evans} II N.~J.,  {Bourke} T.~L.,  {Huard}
  T.~L.,  {Myers} P.~C.,   {Kauffmann} J.,  2008, \mn@doi [\apjs]
  {10.1086/591085}, \href {http://adsabs.harvard.edu/abs/2008ApJS..179..249D}
  {179, 249}

\bibitem[\protect\citeauthoryear{{Eden}, {Moore}, {Plume}  \& {Morgan}}{{Eden}
  et~al.}{2012}]{Eden2012}
{Eden} D.~J.,  {Moore} T.~J.~T.,  {Plume} R.,   {Morgan} L.~K.,  2012, \mn@doi
  [\mnras] {10.1111/j.1365-2966.2012.20840.x}, \href
  {http://adsabs.harvard.edu/abs/2012MNRAS.422.3178E} {422, 3178}

\bibitem[\protect\citeauthoryear{{Eden}, {Moore}, {Morgan}, {Thompson}  \&
  {Urquhart}}{{Eden} et~al.}{2013}]{Eden2013}
{Eden} D.~J.,  {Moore} T.~J.~T.,  {Morgan} L.~K.,  {Thompson} M.~A.,
  {Urquhart} J.~S.,  2013, \mn@doi [\mnras] {10.1093/mnras/stt279}, \href
  {http://adsabs.harvard.edu/abs/2013MNRAS.431.1587E} {431, 1587}

\bibitem[\protect\citeauthoryear{{Eden}, {Moore}, {Urquhart}, {Elia}, {Plume},
  {Rigby}  \& {Thompson}}{{Eden} et~al.}{2015}]{Eden2015}
{Eden} D.~J.,  {Moore} T.~J.~T.,  {Urquhart} J.~S.,  {Elia} D.,  {Plume} R.,
  {Rigby} A.~J.,   {Thompson} M.~A.,  2015, \mn@doi [\mnras]
  {10.1093/mnras/stv1323}, \href
  {http://adsabs.harvard.edu/abs/2015MNRAS.452..289E} {452, 289}

\bibitem[\protect\citeauthoryear{{Egusa}, {Kohno}, {Sofue}, {Nakanishi}  \&
  {Komugi}}{{Egusa} et~al.}{2009}]{Egusa2009}
{Egusa} F.,  {Kohno} K.,  {Sofue} Y.,  {Nakanishi} H.,   {Komugi} S.,  2009,
  \mn@doi [\apj] {10.1088/0004-637X/697/2/1870}, \href
  {http://adsabs.harvard.edu/abs/2009ApJ...697.1870E} {697, 1870}

\bibitem[\protect\citeauthoryear{{Elia} et~al.,}{{Elia}
  et~al.}{2017}]{Elia2017}
{Elia} D.,  et~al., 2017, \mn@doi [\mnras] {10.1093/mnras/stx1357}, \href
  {http://adsabs.harvard.edu/abs/2017MNRAS.471..100E} {471, 100}

\bibitem[\protect\citeauthoryear{{Elmegreen}}{{Elmegreen}}{1980}]{Elmegreen1980}
{Elmegreen} D.~M.,  1980, \mn@doi [\apjs] {10.1086/190666}, \href
  {http://adsabs.harvard.edu/abs/1980ApJS...43...37E} {43, 37}

\bibitem[\protect\citeauthoryear{{Elmegreen} \& {Elmegreen}}{{Elmegreen} \&
  {Elmegreen}}{1986}]{Elmegreen1986}
{Elmegreen} B.~G.,  {Elmegreen} D.~M.,  1986, \mn@doi [\apj] {10.1086/164795},
  \href {http://adsabs.harvard.edu/abs/1986ApJ...311..554E} {311, 554}

\bibitem[\protect\citeauthoryear{{Evans}, {Heiderman}  \&
  {Vutisalchavakul}}{{Evans} et~al.}{2014}]{Evans2014}
{Evans} II N.~J.,  {Heiderman} A.,   {Vutisalchavakul} N.,  2014, \mn@doi
  [\apj] {10.1088/0004-637X/782/2/114}, \href
  {http://adsabs.harvard.edu/abs/2014ApJ...782..114E} {782, 114}

\bibitem[\protect\citeauthoryear{{Foyle}, {Rix}, {Walter}  \& {Leroy}}{{Foyle}
  et~al.}{2010}]{Foyle2010}
{Foyle} K.,  {Rix} H.-W.,  {Walter} F.,   {Leroy} A.~K.,  2010, \mn@doi [\apj]
  {10.1088/0004-637X/725/1/534}, \href
  {http://adsabs.harvard.edu/abs/2010ApJ...725..534F} {725, 534}

\bibitem[\protect\citeauthoryear{{Foyle}, {Rix}, {Dobbs}, {Leroy}  \&
  {Walter}}{{Foyle} et~al.}{2011}]{Foyle2011}
{Foyle} K.,  {Rix} H.-W.,  {Dobbs} C.~L.,  {Leroy} A.~K.,   {Walter} F.,  2011,
  \mn@doi [\apj] {10.1088/0004-637X/735/2/101}, \href
  {http://adsabs.harvard.edu/abs/2011ApJ...735..101F} {735, 101}

\bibitem[\protect\citeauthoryear{{Fumagalli}, {da Silva}  \&
  {Krumholz}}{{Fumagalli} et~al.}{2011}]{Fumigalli2011}
{Fumagalli} M.,  {da Silva} R.~L.,   {Krumholz} M.~R.,  2011, \mn@doi [\apjl]
  {10.1088/2041-8205/741/2/L26}, \href
  {http://adsabs.harvard.edu/abs/2011ApJ...741L..26F} {741, L26}

\bibitem[\protect\citeauthoryear{{Green} et~al.,}{{Green}
  et~al.}{2017}]{Green2017}
{Green} J.~A.,  et~al., 2017, \mn@doi [\mnras] {10.1093/mnras/stx887}, \href
  {http://adsabs.harvard.edu/abs/2017MNRAS.469.1383G} {469, 1383}

\bibitem[\protect\citeauthoryear{{Griffin} et~al.,}{{Griffin}
  et~al.}{2010}]{A&ASpecialIssue-SPIRE}
{Griffin} M.~J.,  et~al., 2010, \mn@doi [\aap] {10.1051/0004-6361/201014519},
  \href {http://adsabs.harvard.edu/abs/2010A%26A...518L...3G} {518, L3+}

\bibitem[\protect\citeauthoryear{{Heiderman}, {Evans}, {Allen}, {Huard}  \&
  {Heyer}}{{Heiderman} et~al.}{2010}]{Heiderman2010}
{Heiderman} A.,  {Evans} II N.~J.,  {Allen} L.~E.,  {Huard} T.,   {Heyer} M.,
  2010, \mn@doi [\apj] {10.1088/0004-637X/723/2/1019}, \href
  {http://adsabs.harvard.edu/abs/2010ApJ...723.1019H} {723, 1019}

\bibitem[\protect\citeauthoryear{{Hermanowicz}, {Kennicutt}  \&
  {Eldridge}}{{Hermanowicz} et~al.}{2013}]{Hermanowicz2013}
{Hermanowicz} M.~T.,  {Kennicutt} R.~C.,   {Eldridge} J.~J.,  2013, \mn@doi
  [\mnras] {10.1093/mnras/stt665}, \href
  {http://adsabs.harvard.edu/abs/2013MNRAS.432.3097H} {432, 3097}

\bibitem[\protect\citeauthoryear{{Heyer} \& {Terebey}}{{Heyer} \&
  {Terebey}}{1998}]{HeyerTerebey1998}
{Heyer} M.~H.,  {Terebey} S.,  1998, \mn@doi [\apj] {10.1086/305881}, \href
  {http://adsabs.harvard.edu/abs/1998ApJ...502..265H} {502, 265}

\bibitem[\protect\citeauthoryear{{Hou} \& {Han}}{{Hou} \&
  {Han}}{2014}]{HouHan2014}
{Hou} L.~G.,  {Han} J.~L.,  2014, \mn@doi [\aap] {10.1051/0004-6361/201424039},
  \href {http://adsabs.harvard.edu/abs/2014A%26A...569A.125H} {569, A125}

\bibitem[\protect\citeauthoryear{{Hou} \& {Han}}{{Hou} \&
  {Han}}{2015}]{HouHan2015}
{Hou} L.~G.,  {Han} J.~L.,  2015, \mn@doi [\mnras] {10.1093/mnras/stv1904},
  \href {http://adsabs.harvard.edu/abs/2015MNRAS.454..626H} {454, 626}

\bibitem[\protect\citeauthoryear{{James} \& {Percival}}{{James} \&
  {Percival}}{2016}]{James2016}
{James} P.~A.,  {Percival} S.~M.,  2016, \mn@doi [\mnras]
  {10.1093/mnras/stv2978}, \href
  {http://adsabs.harvard.edu/abs/2016MNRAS.457..917J} {457, 917}

\bibitem[\protect\citeauthoryear{{Kaufman}, {Bash}, {Hine}, {Rots}, {Elmegreen}
   \& {Hodge}}{{Kaufman} et~al.}{1989}]{Kaufman1989}
{Kaufman} M.,  {Bash} F.~N.,  {Hine} B.,  {Rots} A.~H.,  {Elmegreen} D.~M.,
  {Hodge} P.~W.,  1989, \mn@doi [\apj] {10.1086/167941}, \href
  {http://adsabs.harvard.edu/abs/1989ApJ...345..674K} {345, 674}

\bibitem[\protect\citeauthoryear{{Kennicutt} \& {Evans}}{{Kennicutt} \&
  {Evans}}{2012}]{KennicuttEvans_ARAA}
{Kennicutt} R.~C.,  {Evans} N.~J.,  2012, \mn@doi [\araa]
  {10.1146/annurev-astro-081811-125610}, \href
  {http://adsabs.harvard.edu/abs/2012ARA%26A..50..531K} {50, 531}

\bibitem[\protect\citeauthoryear{{Koda}, {Yagi}, {Boissier}, {Gil de Paz},
  {Imanishi}, {Donovan Meyer}, {Madore}  \& {Thilker}}{{Koda}
  et~al.}{2012}]{Koda2012}
{Koda} J.,  {Yagi} M.,  {Boissier} S.,  {Gil de Paz} A.,  {Imanishi} M.,
  {Donovan Meyer} J.,  {Madore} B.~F.,   {Thilker} D.~A.,  2012, \mn@doi [\apj]
  {10.1088/0004-637X/749/1/20}, \href
  {http://adsabs.harvard.edu/abs/2012ApJ...749...20K} {749, 20}

\bibitem[\protect\citeauthoryear{{Kreckel}, {Blanc}, {Schinnerer}, {Groves},
  {Adamo}, {Hughes}  \& {Meidt}}{{Kreckel} et~al.}{2016}]{Kreckel2016}
{Kreckel} K.,  {Blanc} G.~A.,  {Schinnerer} E.,  {Groves} B.,  {Adamo} A.,
  {Hughes} A.,   {Meidt} S.,  2016, \mn@doi [\apj]
  {10.3847/0004-637X/827/2/103}, \href
  {http://adsabs.harvard.edu/abs/2016ApJ...827..103K} {827, 103}

\bibitem[\protect\citeauthoryear{{Lada}, {Lombardi}  \& {Alves}}{{Lada}
  et~al.}{2010}]{Lada2010}
{Lada} C.~J.,  {Lombardi} M.,   {Alves} J.~F.,  2010, \mn@doi [\apj]
  {10.1088/0004-637X/724/1/687}, \href
  {http://adsabs.harvard.edu/abs/2010ApJ...724..687L} {724, 687}

\bibitem[\protect\citeauthoryear{{Li}, {Gerhard}, {Shen}, {Portail}  \&
  {Wegg}}{{Li} et~al.}{2016}]{Li2016}
{Li} Z.,  {Gerhard} O.,  {Shen} J.,  {Portail} M.,   {Wegg} C.,  2016, \mn@doi
  [\apj] {10.3847/0004-637X/824/1/13}, \href
  {http://adsabs.harvard.edu/abs/2016ApJ...824...13L} {824, 13}

\bibitem[\protect\citeauthoryear{{Lin} \& {Shu}}{{Lin} \&
  {Shu}}{1964}]{LinShu1964}
{Lin} C.~C.,  {Shu} F.~H.,  1964, \mn@doi [\apj] {10.1086/147955}, \href
  {http://adsabs.harvard.edu/abs/1964ApJ...140..646L} {140, 646}

\bibitem[\protect\citeauthoryear{{Lindblad}}{{Lindblad}}{1960}]{Lindblad1960}
{Lindblad} P.~O.,  1960, Stockholms Observatoriums Annaler, \href
  {http://adsabs.harvard.edu/abs/1960StoAn..21....4L} {21}

\bibitem[\protect\citeauthoryear{{Lombardi}, {Lada}  \& {Alves}}{{Lombardi}
  et~al.}{2013}]{Lombardi2013}
{Lombardi} M.,  {Lada} C.~J.,   {Alves} J.,  2013, \mn@doi [\aap]
  {10.1051/0004-6361/201321827}, \href
  {http://adsabs.harvard.edu/abs/2013A%26A...559A..90L} {559, A90}

\bibitem[\protect\citeauthoryear{{Louie}, {Koda}  \& {Egusa}}{{Louie}
  et~al.}{2013}]{Louie2013}
{Louie} M.,  {Koda} J.,   {Egusa} F.,  2013, \mn@doi [\apj]
  {10.1088/0004-637X/763/2/94}, \href
  {http://adsabs.harvard.edu/abs/2013ApJ...763...94L} {763, 94}

\bibitem[\protect\citeauthoryear{{Molinari}, {Pezzuto}, {Cesaroni}, {Brand},
  {Faustini}  \& {Testi}}{{Molinari} et~al.}{2008}]{Molinari2008}
{Molinari} S.,  {Pezzuto} S.,  {Cesaroni} R.,  {Brand} J.,  {Faustini} F.,
  {Testi} L.,  2008, \mn@doi [\aap] {10.1051/0004-6361:20078661}, \href
  {http://adsabs.harvard.edu/abs/2008A%26A...481..345M} {481, 345}

\bibitem[\protect\citeauthoryear{{Molinari} et~al.,}{{Molinari}
  et~al.}{2010a}]{Molinari2010a}
{Molinari} S.,  et~al., 2010a, \mn@doi [\pasp] {10.1086/651314}, \href
  {http://adsabs.harvard.edu/abs/2010PASP..122..314M} {122, 314}

\bibitem[\protect\citeauthoryear{{Molinari} et~al.,}{{Molinari}
  et~al.}{2010b}]{Molinari2010b}
{Molinari} S.,  et~al., 2010b, \mn@doi [\aap] {10.1051/0004-6361/201014659},
  \href {http://adsabs.harvard.edu/abs/2010A%26A...518L.100M} {518, L100}

\bibitem[\protect\citeauthoryear{{Molinari}, {Merello}, {Elia}, {Cesaroni},
  {Testi}  \& {Robitaille}}{{Molinari} et~al.}{2016}]{Molinari2016c}
{Molinari} S.,  {Merello} M.,  {Elia} D.,  {Cesaroni} R.,  {Testi} L.,
  {Robitaille} T.,  2016, \mn@doi [\apjl] {10.3847/2041-8205/826/1/L8}, \href
  {http://adsabs.harvard.edu/abs/2016ApJ...826L...8M} {826, L8}

\bibitem[\protect\citeauthoryear{{Moore}, {Urquhart}, {Morgan}  \&
  {Thompson}}{{Moore} et~al.}{2012}]{Moore2012}
{Moore} T.~J.~T.,  {Urquhart} J.~S.,  {Morgan} L.~K.,   {Thompson} M.~A.,
  2012, \mn@doi [\mnras] {10.1111/j.1365-2966.2012.21740.x}, \href
  {http://adsabs.harvard.edu/abs/2012MNRAS.426..701M} {426, 701}

\bibitem[\protect\citeauthoryear{{Mottram} et~al.,}{{Mottram}
  et~al.}{2011}]{Mottram2011}
{Mottram} J.~C.,  et~al., 2011, \mn@doi [\apjl] {10.1088/2041-8205/730/2/L33},
  \href {http://adsabs.harvard.edu/abs/2011ApJ...730L..33M} {730, L33}

\bibitem[\protect\citeauthoryear{{Myers} \& {Ladd}}{{Myers} \&
  {Ladd}}{1993}]{Myers1993}
{Myers} P.~C.,  {Ladd} E.~F.,  1993, \mn@doi [\apjl] {10.1086/186956}, \href
  {http://adsabs.harvard.edu/abs/1993ApJ...413L..47M} {413, L47}

\bibitem[\protect\citeauthoryear{{Pflamm-Altenburg}, {Weidner}  \&
  {Kroupa}}{{Pflamm-Altenburg} et~al.}{2007}]{Pflamm-Altenburg2007}
{Pflamm-Altenburg} J.,  {Weidner} C.,   {Kroupa} P.,  2007, \mn@doi [\apj]
  {10.1086/523033}, \href {http://adsabs.harvard.edu/abs/2007ApJ...671.1550P}
  {671, 1550}

\bibitem[\protect\citeauthoryear{{Pflamm-Altenburg},
  {Gonz{\'a}lez-L{\'o}pezlira}  \& {Kroupa}}{{Pflamm-Altenburg}
  et~al.}{2013}]{Pflamm-Altenburg2013}
{Pflamm-Altenburg} J.,  {Gonz{\'a}lez-L{\'o}pezlira} R.~A.,   {Kroupa} P.,
  2013, \mn@doi [\mnras] {10.1093/mnras/stt1474}, \href
  {http://adsabs.harvard.edu/abs/2013MNRAS.435.2604P} {435, 2604}

\bibitem[\protect\citeauthoryear{{Poglitsch} et~al.,}{{Poglitsch}
  et~al.}{2010}]{A&ASpecialIssue-PACS}
{Poglitsch} A.,  et~al., 2010, \mn@doi [\aap] {10.1051/0004-6361/201014535},
  \href {http://adsabs.harvard.edu/abs/2010A%26A...518L...2P} {518, L2+}

\bibitem[\protect\citeauthoryear{{Ragan} et~al.,}{{Ragan}
  et~al.}{2012}]{Ragan2012b}
{Ragan} S.~E.,  et~al., 2012, \mn@doi [\aap] {10.1051/0004-6361/201219232},
  \href {http://adsabs.harvard.edu/abs/2012A%26A...547A..49R} {547, A49}

\bibitem[\protect\citeauthoryear{{Ragan}, {Moore}, {Eden}, {Hoare}, {Elia}  \&
  {Molinari}}{{Ragan} et~al.}{2016}]{Ragan2016}
{Ragan} S.~E.,  {Moore} T.~J.~T.,  {Eden} D.~J.,  {Hoare} M.~G.,  {Elia} D.,
  {Molinari} S.,  2016, \mn@doi [\mnras] {10.1093/mnras/stw1870}, \href
  {http://adsabs.harvard.edu/abs/2016MNRAS.462.3123R} {462, 3123}

\bibitem[\protect\citeauthoryear{{Rebolledo}, {Wong}, {Xue}, {Leroy}, {Koda}
  \& {Donovan Meyer}}{{Rebolledo} et~al.}{2015}]{Rebolledo2015}
{Rebolledo} D.,  {Wong} T.,  {Xue} R.,  {Leroy} A.,  {Koda} J.,   {Donovan
  Meyer} J.,  2015, \mn@doi [\apj] {10.1088/0004-637X/808/1/99}, \href
  {http://adsabs.harvard.edu/abs/2015ApJ...808...99R} {808, 99}

\bibitem[\protect\citeauthoryear{{Reed}}{{Reed}}{2000}]{Reed2000}
{Reed} B.~C.,  2000, \mn@doi [\aj] {10.1086/301421}, \href
  {http://adsabs.harvard.edu/abs/2000AJ....120..314R} {120, 314}

\bibitem[\protect\citeauthoryear{{Reid} et~al.,}{{Reid}
  et~al.}{2014}]{Reid2014a}
{Reid} M.~J.,  et~al., 2014, \mn@doi [\apj] {10.1088/0004-637X/783/2/130},
  \href {http://adsabs.harvard.edu/abs/2014ApJ...783..130R} {783, 130}

\bibitem[\protect\citeauthoryear{{Reid}, {Dame}, {Menten}  \&
  {Brunthaler}}{{Reid} et~al.}{2016}]{Reid2016}
{Reid} M.~J.,  {Dame} T.~M.,  {Menten} K.~M.,   {Brunthaler} A.,  2016, \mn@doi
  [\apj] {10.3847/0004-637X/823/2/77}, \href
  {http://adsabs.harvard.edu/abs/2016ApJ...823...77R} {823, 77}

\bibitem[\protect\citeauthoryear{{Roberts}}{{Roberts}}{1969}]{Roberts1969}
{Roberts} W.~W.,  1969, \mn@doi [\apj] {10.1086/150177}, \href
  {http://adsabs.harvard.edu/abs/1969ApJ...158..123R} {158, 123}

\bibitem[\protect\citeauthoryear{{Robitaille}, {Churchwell}, {Benjamin},
  {Whitney}, {Wood}, {Babler}  \& {Meade}}{{Robitaille}
  et~al.}{2012}]{Robitaille2012}
{Robitaille} T.~P.,  {Churchwell} E.,  {Benjamin} R.~A.,  {Whitney} B.~A.,
  {Wood} K.,  {Babler} B.~L.,   {Meade} M.~R.,  2012, \mn@doi [\aap]
  {10.1051/0004-6361/201219073}, \href
  {http://adsabs.harvard.edu/abs/2012A%26A...545A..39R} {545, A39}

\bibitem[\protect\citeauthoryear{{Roman-Duval}, {Jackson}, {Heyer}, {Johnson},
  {Rathborne}, {Shah}  \& {Simon}}{{Roman-Duval}
  et~al.}{2009}]{Roman-Duval2009}
{Roman-Duval} J.,  {Jackson} J.~M.,  {Heyer} M.,  {Johnson} A.,  {Rathborne}
  J.,  {Shah} R.,   {Simon} R.,  2009, \mn@doi [\apj]
  {10.1088/0004-637X/699/2/1153}, \href
  {http://adsabs.harvard.edu/abs/2009ApJ...699.1153R} {699, 1153}

\bibitem[\protect\citeauthoryear{{Roman-Duval}, {Jackson}, {Heyer}, {Rathborne}
   \& {Simon}}{{Roman-Duval} et~al.}{2010}]{Roman-Duval2010}
{Roman-Duval} J.,  {Jackson} J.~M.,  {Heyer} M.,  {Rathborne} J.,   {Simon} R.,
   2010, \mn@doi [\apj] {10.1088/0004-637X/723/1/492}, \href
  {http://adsabs.harvard.edu/abs/2010ApJ...723..492R} {723, 492}

\bibitem[\protect\citeauthoryear{{Russeil} et~al.,}{{Russeil}
  et~al.}{2011}]{Russeil2011}
{Russeil} D.,  et~al., 2011, \mn@doi [\aap] {10.1051/0004-6361/201015852},
  \href {http://adsabs.harvard.edu/abs/2011A%26A...526A.151R} {526, A151}

\bibitem[\protect\citeauthoryear{{Sawada}, {Hasegawa}, {Sugimoto}, {Koda}  \&
  {Handa}}{{Sawada} et~al.}{2012a}]{Sawada2012a}
{Sawada} T.,  {Hasegawa} T.,  {Sugimoto} M.,  {Koda} J.,   {Handa} T.,  2012a,
  \mn@doi [\apj] {10.1088/0004-637X/752/2/118}, \href
  {http://adsabs.harvard.edu/abs/2012ApJ...752..118S} {752, 118}

\bibitem[\protect\citeauthoryear{{Sawada}, {Hasegawa}  \& {Koda}}{{Sawada}
  et~al.}{2012b}]{Sawada2012b}
{Sawada} T.,  {Hasegawa} T.,   {Koda} J.,  2012b, \mn@doi [\apjl]
  {10.1088/2041-8205/759/1/L26}, \href
  {http://adsabs.harvard.edu/abs/2012ApJ...759L..26S} {759, L26}

\bibitem[\protect\citeauthoryear{{Schinnerer} et~al.,}{{Schinnerer}
  et~al.}{2013}]{Schinnerer2013}
{Schinnerer} E.,  et~al., 2013, \mn@doi [\apj] {10.1088/0004-637X/779/1/42},
  \href {http://adsabs.harvard.edu/abs/2013ApJ...779...42S} {779, 42}

\bibitem[\protect\citeauthoryear{{Schinnerer} et~al.,}{{Schinnerer}
  et~al.}{2017}]{Schinnerer2017}
{Schinnerer} E.,  et~al., 2017, \mn@doi [\apj] {10.3847/1538-4357/836/1/62},
  \href {http://adsabs.harvard.edu/abs/2017ApJ...836...62S} {836, 62}

\bibitem[\protect\citeauthoryear{{Sormani}, {Binney}  \& {Magorrian}}{{Sormani}
  et~al.}{2015}]{Sormani2015}
{Sormani} M.~C.,  {Binney} J.,   {Magorrian} J.,  2015, \mn@doi [\mnras]
  {10.1093/mnras/stv2067}, \href
  {http://adsabs.harvard.edu/abs/2015MNRAS.454.1818S} {454, 1818}

\bibitem[\protect\citeauthoryear{{Tamburro}, {Rix}, {Walter}, {Brinks}, {de
  Blok}, {Kennicutt}  \& {Mac Low}}{{Tamburro} et~al.}{2008}]{Tamburro2008}
{Tamburro} D.,  {Rix} H.-W.,  {Walter} F.,  {Brinks} E.,  {de Blok} W.~J.~G.,
  {Kennicutt} R.~C.,   {Mac Low} M.-M.,  2008, \mn@doi [\aj]
  {10.1088/0004-6256/136/6/2872}, \href
  {http://adsabs.harvard.edu/abs/2008AJ....136.2872T} {136, 2872}

\bibitem[\protect\citeauthoryear{{Tenjes}, {Tuvikene}, {Tamm}, {Kipper}  \&
  {Tempel}}{{Tenjes} et~al.}{2017}]{Tenjes2017}
{Tenjes} P.,  {Tuvikene} T.,  {Tamm} A.,  {Kipper} R.,   {Tempel} E.,  2017,
  \mn@doi [\aap] {10.1051/0004-6361/201629991}, \href
  {http://adsabs.harvard.edu/abs/2017A%26A...600A..34T} {600, A34}

\bibitem[\protect\citeauthoryear{{Toomre}}{{Toomre}}{1969}]{Toomre1969}
{Toomre} A.,  1969, \mn@doi [\apj] {10.1086/150250}, \href
  {http://adsabs.harvard.edu/abs/1969ApJ...158..899T} {158, 899}

\bibitem[\protect\citeauthoryear{{Urquhart} et~al.,}{{Urquhart}
  et~al.}{2012}]{Urquhart2012}
{Urquhart} J.~S.,  et~al., 2012, \mn@doi [\mnras]
  {10.1111/j.1365-2966.2011.20157.x}, \href
  {http://adsabs.harvard.edu/abs/2012MNRAS.420.1656U} {420, 1656}

\bibitem[\protect\citeauthoryear{{Urquhart} et~al.,}{{Urquhart}
  et~al.}{2013}]{Urquhart2013}
{Urquhart} J.~S.,  et~al., 2013, \mn@doi [\mnras] {10.1093/mnras/stt1310},
  \href {http://adsabs.harvard.edu/abs/2013MNRAS.435..400U} {435, 400}

\bibitem[\protect\citeauthoryear{{Urquhart}, {Figura}, {Moore}, {Hoare},
  {Lumsden}, {Mottram}, {Thompson}  \& {Oudmaijer}}{{Urquhart}
  et~al.}{2014a}]{Urquhart2014a}
{Urquhart} J.~S.,  {Figura} C.~C.,  {Moore} T.~J.~T.,  {Hoare} M.~G.,
  {Lumsden} S.~L.,  {Mottram} J.~C.,  {Thompson} M.~A.,   {Oudmaijer} R.~D.,
  2014a, \mn@doi [\mnras] {10.1093/mnras/stt2006}, \href
  {http://adsabs.harvard.edu/abs/2014MNRAS.437.1791U} {437, 1791}

\bibitem[\protect\citeauthoryear{{Urquhart} et~al.,}{{Urquhart}
  et~al.}{2014b}]{Urquhart2014c}
{Urquhart} J.~S.,  et~al., 2014b, \mn@doi [\mnras] {10.1093/mnras/stu1207},
  \href {http://adsabs.harvard.edu/abs/2014MNRAS.443.1555U} {443, 1555}

\bibitem[\protect\citeauthoryear{{Urquhart} et~al.,}{{Urquhart}
  et~al.}{2018}]{Urquhart2017}
{Urquhart} J.~S.,  et~al., 2018, \mn@doi [\mnras] {10.1093/mnras/stx2258},
  \href {http://adsabs.harvard.edu/abs/2018MNRAS.473.1059U} {473, 1059}

\bibitem[\protect\citeauthoryear{{Vall{\'e}e}}{{Vall{\'e}e}}{1995}]{Vallee1995}
{Vall{\'e}e} J.~P.,  1995, \mn@doi [\apj] {10.1086/176470}, \href
  {http://adsabs.harvard.edu/abs/1995ApJ...454..119V} {454, 119}

\bibitem[\protect\citeauthoryear{{Vall{\'e}e}}{{Vall{\'e}e}}{2014a}]{Vallee2014a}
{Vall{\'e}e} J.~P.,  2014a, \mn@doi [\aj] {10.1088/0004-6256/148/1/5}, \href
  {http://adsabs.harvard.edu/abs/2014AJ....148....5V} {148, 5}

\bibitem[\protect\citeauthoryear{{Vall{\'e}e}}{{Vall{\'e}e}}{2014b}]{Vallee2014c}
{Vall{\'e}e} J.~P.,  2014b, \mn@doi [\apjs] {10.1088/0067-0049/215/1/1}, \href
  {http://adsabs.harvard.edu/abs/2014ApJS..215....1V} {215, 1}

\bibitem[\protect\citeauthoryear{{Vall{\'e}e}}{{Vall{\'e}e}}{2016}]{Vallee2016b}
{Vall{\'e}e} J.~P.,  2016, \mn@doi [\apj] {10.3847/0004-637X/821/1/53}, \href
  {http://adsabs.harvard.edu/abs/2016ApJ...821...53V} {821, 53}

\bibitem[\protect\citeauthoryear{{Vogel}, {Kulkarni}  \& {Scoville}}{{Vogel}
  et~al.}{1988}]{Vogel1988}
{Vogel} S.~N.,  {Kulkarni} S.~R.,   {Scoville} N.~Z.,  1988, \mn@doi [\nat]
  {10.1038/334402a0}, \href {http://adsabs.harvard.edu/abs/1988Natur.334..402V}
  {334, 402}

\bibitem[\protect\citeauthoryear{{Vutisalchavakul} \&
  {Evans}}{{Vutisalchavakul} \& {Evans}}{2013}]{Vuti2013}
{Vutisalchavakul} N.,  {Evans} II N.~J.,  2013, \mn@doi [\apj]
  {10.1088/0004-637X/765/2/129}, \href
  {http://adsabs.harvard.edu/abs/2013ApJ...765..129V} {765, 129}

\makeatother
\end{thebibliography}
